\documentclass[conference]{IEEEtran}
\usepackage{cite}
\usepackage{amsmath,amssymb,amsfonts}
\usepackage{algorithmic}
\usepackage{graphicx}
\usepackage{textcomp}
\usepackage{xcolor}
\usepackage[hyphens]{url}
\usepackage{fancyhdr}
\usepackage{hyperref}
\usepackage{amsmath}
\usepackage[braket, qm]{qcircuit}
\usepackage{array}
\usepackage{soul}

\usepackage[strict]{changepage}

\usepackage[most]{tcolorbox}
\newcounter{obscounter}
\renewcommand{\theobscounter}{\Roman{obscounter}}
\definecolor{custompurple}{HTML}{53257F}
\newtcolorbox{ObservationBox}[2][]{text width=0.94\linewidth,
colbacktitle=custompurple,enhanced,
attach boxed title to top left={yshift=-2mm,xshift=3mm},
boxed title style={sharp corners},top=6pt,bottom=2pt,
title=#2,colback=custompurple!10!white, left=4pt, right=2pt}
\newcommand{\obs}[1]{
\refstepcounter{obscounter}
\begin{ObservationBox}{\textbf{Observation \theobscounter}}
\par\noindent
#1
\end{ObservationBox}}

\begin{document}

%%%%%%%%%%%%%%%%%%%%%%%%%%%%%%%%%%%%%%%%
%%%%%%%%%%%%%% -- UPDATE -- %%%%%%%%%%%%%%%
\title{State of practice: evaluating GPU performance of state vector and tensor network methods}
%%%%%%%%%%%%%%%%%%%%%%%%%%%%%%%%%%%%%%%%

\author{\IEEEauthorblockN{Marzio Vallero}
\IEEEauthorblockA{
\textit{University of Trento}\\
Trento, Italy \\
marzio.vallero@unitn.it}
\and
\IEEEauthorblockN{Paolo Rech}
\IEEEauthorblockA{
\textit{University of Trento}\\
Trento, Italy \\
paolo.rech@unitn.it}
\and
\IEEEauthorblockN{Flavio Vella}
\IEEEauthorblockA{
\textit{University of Trento}\\
Trento, Italy \\
flavio.vella@unitn.it}
}

\maketitle

%%%%%%%%%%%%%%%%%%%%%%%%%%%%%%%%%%%%%%%%
%%%%%%%% -- PAPER CONTENT STARTS -- %%%%%%%%%

\begin{abstract}

The frontier of quantum computing (QC) simulation on classical hardware is quickly reaching the hard scalability limits for computational feasibility.
Nonetheless, there is still a need to simulate large quantum systems classically, as the Noisy Intermediate Scale Quantum (NISQ) devices are yet to be considered fault tolerant and performant enough in terms of operations per second.
Each of the two main exact simulation techniques, state vector and tensor network simulators, boasts specific limitations.
% The exponential memory requirement of state vector simulation, when compared to the qubit register sizes of currently available quantum computers, quickly saturates the capacity of the top HPC machines.
% Tensor network contraction approaches, which encode quantum circuits into tensor networks and then contract them over an output bit string to obtain its probability amplitude, still fall short of the inherent complexity of finding an optimal contraction path, which maps to a max-cut problem on a dense mesh, a notably NP-hard problem. 

This article investigates the limits of current state-of-the-art simulation techniques on a test bench made of eight widely used quantum subroutines, each in different configurations, with a special emphasis on performance.
We perform both single process and distributed scaleability experiments on a supercomputer.  
%State vector simulations and tensor network contractions are performed through the \textit{cuQuantum} library. %%PR: non nell'abstract
We correlate the performance measures from such experiments with the metrics that characterise the benchmark circuits, identifying the main reasons behind the observed performance trends.
%OLD1: Based on our observations, we highlight how the structure of a quantum circuit can lead to a speedup of up to an order of magnitude improvement with tensor network contraction when compared to state vector simulators.
%OLD2: From our observations, we highlight how, given the structure of a quantum circuit, we can select the best simulation strategy, obtaining a speedup of up to an order of magnitude.
Specifically, we perform distributed sliced tensor contractions, and we analyse the impact of pathfinding quality on contraction time, correlating both results with topological circuit characteristics.
From our observations, given the structure of a quantum circuit and the number of qubits, we highlight how to select the best simulation strategy, demonstrating how preventive circuit analysis can guide and improve simulation performance by more than an order of magnitude.

% \todo[inline]{summarize results and impact here.}


\end{abstract}

\section{Introduction}
\label{intro}
% What is quantum computing and what can it be used for?
% Do we really need quantum computers, or can we just simulate?
% Applications of simulation for verifying the results of real quantum hardware
% Small touch on the issues of simulation (memory occupancy)
Do we really need to build quantum computers?\newline
The answer is blatantly affirmative, however knowing where the hard boundary for the classical simulation of quantum systems lies is most definitely \textit{not} a straightforward task.

The quantum computing paradigm, since its theoretical inception by one of the forefathers of modern physics R. Feynman in 1982\cite{feynman1982}, served to extend the classical definition of computation in order to better describe the quantum properties of nature, under the hypothesis that \textit{an experiment}, which is a purposefully built physical system, can be said to be \textit{performing computations} under specific conditions\cite{horsman2014}. Years later, many quantum algorithms would be proposed, able to exploit the characteristics of this new paradigm to achieve unprecedented speedups in unstructured search problems \cite{grover1996} or to efficiently solve hard mathematical problems, such as prime factorization of large numbers \cite{shor1997}, and other more recent applications in the fields of optimisation \cite{qaoa_ref}, machine learning \cite{vqe_ref} and chemistry \cite{quantum-drug}.\\
In the last decade, progress in the various technologies used for building quantum devices has reached commercial applications, such as the cloud services proposed by IBM, D-Wave and Google, among others. These systems, however, are yet to be considered mature, as most still fail in preserving coherent quantum states, suffer cross-talk among their constituent elements and employ imperfect logical operators. Furthermore, supercondution-based devices, which make up a wide margin of the operational quantum devices as of the writing of this article, have been recently proven to be exceedingly susceptible to external radiation events \cite{vepsalainen2020, wilen2021, mcewen2021}, once again hindering their applicability.\\
Meanwhile, the top performing supercomputers have surpassed the exascale number of floating operations per second \cite{TOP5002023}. Naturally, such an achievement pushes up the threshold for the complexity of quantum systems that can be classically simulated over a reasonable time span \cite{hoefler23}. This opens up the possibility for some quantum algorithms to be efficiently converted into quantum-inspired classical algorithms. Moreover, the testing and validation of newly proposed quantum algorithms, and the accuracy measurement of real quantum devices' outputs must be done through simulation, to double check results and ensure correct operation. With the increasingly higher optimisation of specialised software libraries over commodity hardware accelerators, such as graphic processing units (GPUs), the rapid simulation of \textit{small} quantum algorithms is becoming more and more easy to accomplish \cite{cuquantum,qsim,qiskit,pennylane}. Such practices are, of course, still limited in memory and time by the exponential requirements of quantum simulation, but better exploitation of current classical computation resources may lead to efficient simulation of small quantum subroutines without needing to compensate for hardware-level noise and external radiation events, possibly reducing or annulling the queries made to cloud-based quantum computers for low qubit sized problems.  

The purpose of this article is to understand where the limit for efficient quantum simulation on classical hardware lies, emphasising the computational aspects, such as distributed performance, scaleability, time and memory footprints of quantum algorithms, with the objective to find quantum circuit features that correlate to simulation performance.
There are various classical simulation techniques for quantum algorithms other than state vector simulation and tensor network contractions, such as stabilisers theory or p-block simulation. Notably, however, stabiliser theory is limited to the application of a non-universal set of Clifford gates, whilst p-block simulation leverages approximations in the representation that lead to very limited entanglement.
Being the only two approaches able to find an exact solution, the analysis we propose focuses on state vector simulation and tensor network contraction techniques.

% \todo[inline]{There are two (three??) major ways to simulate a quantum circuit in a classical machine: X, Y, Z?, each with specific benefits and limitations. }
% \todo[inline]{Our scope is to investigate possible bottlenecks, inefficiencies, and intrinsic limits to identify the most efficient simlation platform for a given quantum algorithm.}

The questions seek an answer for are: what is the performance of state of the art quantum simulation methods? Which topological features of a quantum circuit correlate to simulation performance, and which simulation approach is more suitable? Are there limitations to distributed quantum simulation, and can we predict them?

We prove that, by profiling quantum circuits with the approach presented in this paper, the simulation time can reach a speedup of \textit{up to one order of magnitude}, especially for large quantum circuits, on a single GPU.
Furthermore, we report the results from distributed tensor contraction simulations, highlighting speedups of more than $364\times$ with respect to single GPU performance, and the we trace the impact of pathfinding quality on the contraction performance, obtaining speedups of up to $4.79\times$ through tuning.
The proposed circuit metrics to performance correlation is achieved by characterising a purposefully selected suite of well known quantum circuit subroutines according to objective metrics, and checking how those scale with respect to the size of the quantum circuit. All the circuits we consider are parameterisable over the number of qubits in the system, and some of them feature additional customisation parameters, such as layer repetition. Moreover, they have been selected as to have practical applicability in terms of exact simulation. %: for this reason, error correcting codes or circuits such as the Greenberger–Horne–Zeilinger have been avoided.
These same circuits have been simulated on CINECA's Leonardo supercomputer, the $7^{th}$ supercomputer in the Top500 list, using both state-vector and tensor network contraction methods through NVIDIA's \textit{cuQuantum} library \cite{cuquantum}, highlighting which one boasts the better performance for each workload. Our work proposes a practical methodology to pick the most efficient simulation strategy according to a given set of static characteristics of the circuit.

The rest of the paper is organized as follows:
Section \ref{background_sec} provides a short summary of quantum computing as a whole, and it is followed by Section \ref{simulation_sec}, which introduces how classical algorithms for quantum simulation work. Section \ref{benchmarks_metrics_sec} gives a definition of the metrics and of the quantum circuits considered for this study. Section \ref{results_sec} characterises the quantum circuits according to the aforementioned metrics, then presents performance results with respect to execution times and peak memory occupancy, scaling of distributed tensor network contractions and impact of pathfinding resources on tensor network contraction times. Lastly, Section \ref{conslusions_sec} concludes the paper by expanding on the hereby presented work by opening new paths for investigation in future works.

\section{Background}
\label{background_sec}
Nature, on an atomic and subatomic scale, is inherently quantum. When tasked with modeling and simulating the properties of such atomic-scale phenomena, it is reasonable to do so with objects that are able to express the same quantum properties that are to be investigated. This implies that the binary computation paradigm has to evolve towards a representation which encompasses such additional characteristics.

\subsection{Quantum Computing}
% What did Feynman say about quantum?
% What is a qubit? What are its properties?
 Quantum computing is an expansion of binary computing able to tackle any problem that the latter approach can tackle, whilst at the same time providing more efficient solutions to problems that are deemed intractable in the classical domain. This is achieved by exploiting \textit{ad hoc} resources, namely \textit{superposition} and \textit{entanglement}.\\ Quantum computers make use of \textit{qubits}, the quantum counterpart of classical bits, to encode information. Each qubit is described by two complex probability amplitudes, as follows
 \begin{equation}
 \label{bell_state}
     \ket{\psi} = \alpha \ket{0} + \beta \ket{1}, \quad \alpha,\beta \in \mathbb{C}
 \end{equation}
 
Each amplitude, when squared, represents the probability for the qubit to collapse on its corresponding classical state when observed. It must hold for the sum of the two squared probabilities to equal unity, i.e. $\alpha^2 + \beta^2 = 1$.
Whenever the two amplitudes assume non-integer values, the qubit is said to be in \textit{superposition} between the two basis states. The property of \textit{entanglement} refers to the fact that two or more qubits can share a non-classical correlation, such that when one of the entangled qubits is observed, it lets us infer information regarding the state of the others without measuring them. This second property sprouted issues with locality of quantum mechanics in the well-known EPR paradox \cite{epr_paradox}, later solved by J.S. Bell \cite{bell_debunks_epr}, which defined the minimum set of these namesake entangled quantum states. These two qubit  states cannot be described as the product of two independent qubits, such as
\begin{equation}
    \ket{\Phi^+} = \alpha \ket{00} + \beta \ket{11} 
\end{equation}

Entanglement is believed to be the fundamental resource responsible for quantum speedup, although superposition plays an important role as well, since circuits with low entanglement have been proven to be trivial to simulate \cite{Vidal2003}.

\subsection{Quantum circuits}
% We can use a system of "gates" to encode quantum operators, the basic blocks for building quantum programs
% Quantum circuits are read from left to right, following the flow of information
% Gates encode matrices that modify the system's wavefunction, either on singular qubits or on multiple qubits at a time
% Clifford gates and universal gate set
Algorithms in the QC field are expressed via quantum circuits, a graphical notation derived from Penrose's notation\cite{Coecke2011}. They are read from left to right, following the flow of information. Operations on one or multiple qubits are applied via \textit{quantum gates}, which are represented by $2^N \times 2^N$ unitary matrices, with $N$ being the number of qubits acted upon by the gate. The most basic single qubit operators include the Pauli X, Y, Z gates and the basis-swap Hadamard gate, represented by the following matrices
\begin{equation}
    \begin{split}
        X = \begin{bmatrix}
            0 & 1 \\ 1 & 0
        \end{bmatrix}
        , & \quad
        Y = \begin{bmatrix}
            0 & -i \\ i & 0
        \end{bmatrix}
        \\
        Z = \begin{bmatrix}
            1 & 0 \\ 0 & -1
        \end{bmatrix}
        , & \quad 
        H = \frac{1}{\sqrt{2}} \begin{bmatrix}
             1 & 1 \\ 1 & -1
        \end{bmatrix}
    \end{split}
\end{equation}

Multiple qubit gates are generally employed to generate entanglement, such as the controlled-not (CNOT) gate, that applies an X gate to a target qubit if the control qubit is in the $\ket{1}$ state. The matrix representation of the CNOT gate is 
\begin{equation}
    CNOT = \begin{bmatrix}
        1 & 0 & 0 & 0 \\ 0 & 1 & 0 & 0 \\ 0 & 0 & 0 & 1 \\ 0 & 0 & 1 & 0
    \end{bmatrix}
\end{equation}

Similarly to the binary paradigm, it is possible to define universal quantum gate sets that can be used to encode any operator as the composition of these fundamental ones. This is at the basis of the construction of universal quantum computers.

\begin{figure}[!ht]
\begin{center}
    \scalebox{1.5}{
        \Qcircuit @C=1.0em @R=0.2em @!R {
        	 	\nghost{\ket{0} :  } & \lstick{\ket{0} :  } & \gate{\mathrm{H}} & \ctrl{1} & \qw & \meter & \qw & \qw \\
        	 	\nghost{\ket{0} :  } & \lstick{\ket{0} :  } & \qw & \targ & \qw & \qw & \meter & \qw \\
        	 	\nghost{\mathrm{{c} :  }} & \lstick{\mathrm{{c[2]} :  }} & \lstick{/_{_{2}}} \cw & \cw & \cw & \dstick{_{_{\hspace{0.0em}0}}} \cw \ar @{<=} [-2,0] & \dstick{_{_{\hspace{0.0em}1}}} \cw \ar @{<=} [-1,0] & \cw \\
        }
    }

        \caption{Quantum circuit encoding the $\ket{\Phi^+}$ Bell state and measure operators.}
        \label{bell_circuit}
\end{center}
\end{figure}
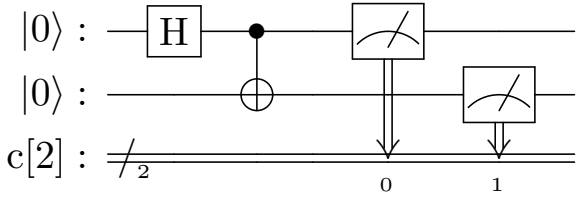

Figure \ref{bell_circuit} shows the quantum circuit that encodes the $\ket{\Phi^+}$ Bell state of Equation \ref{bell_state}: the two qubits are initialised in state $\ket{0}$, then the fist qubit is put in the equiprobable superposition state $\ket{+}$ and is used as the control of a CNOT gate. The second qubit is now entangled the first.

\begin{figure*}[!ht]
    \centering
    \includegraphics[width=.49\linewidth]{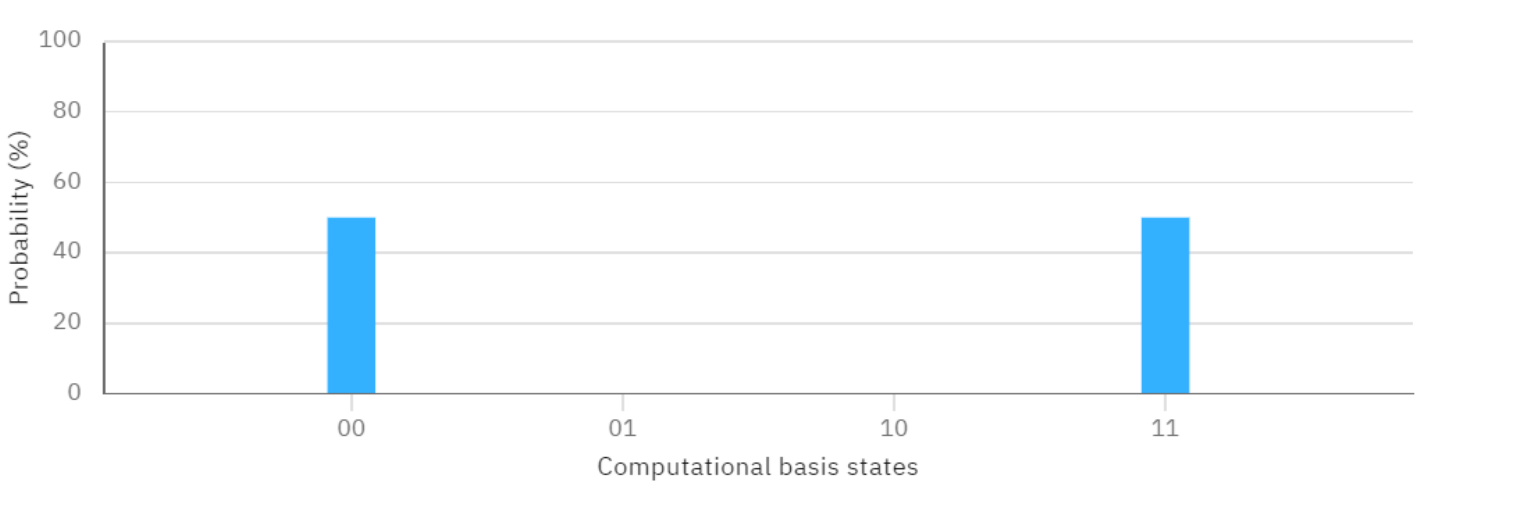}
    \includegraphics[width=.49\linewidth]{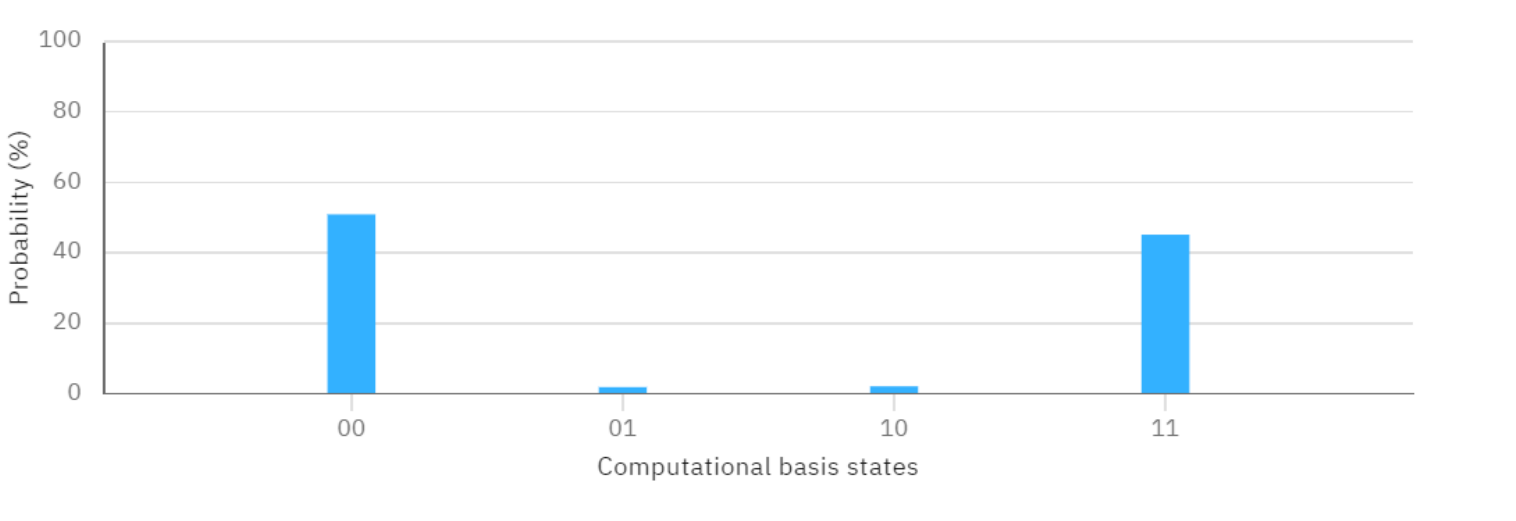}
    
    \caption{Comparison between the theoretical output distribution of the Bell circuit (left) and the experimental output distribution obtained from a real quantum device, an IBM Falcon r4T processor, over 1024 measurement shots (right).}
    \label{compare_sim_device}
    
\end{figure*}

The execution of a quantum circuit on a real quantum device yields a single \textit{classical} output bit string. By repeatedly executing and measuring the same circuit, it is possible to sample an output distribution, which converges to the element-wise square of the state vector made of the probability amplitudes of each possible quantum state in the system. Experimental results and expected theoretical results oftentimes differ, due to intrinsic errors in the quantum device, cross talk among qubits and the number of samples made. Such an example has been provided in Figure \ref{compare_sim_device}.
Classical simulation of quantum systems is mainly used to compute the expected theoretical output distribution of a quantum circuit, so as to later compare it to the output of a real quantum device.

\section{Quantum circuit simulation}
\label{simulation_sec}
% If we could simulate all quantum circuits, we wouldn't need to build real quantum computers, but this is impossible
% The need for validation of quantum devices requires efficient quantum simulation, to help testing new algorithms and calibrate new quantum devices
% There are two main techniques for simulation, state vectors and tensor network contractions
% \todo[inline]{Add list of things that limit simulation capabilities}
Quantum simulation can refer to two concepts: either the usage of real quantum devices to simulate other quantum systems, or the usage of classical machines to compute the theoretical output of a quantum algorithm. For the sake of clarity, this paper will always refer to the latter when talking about \textit{quantum circuit simulation}.
The objective of simulation is not to thwart the development of real quantum devices, but rather to validate the outputs of such machines against their theoretical expected outputs. Moreover, given the still relatively scarce availability of real quantum devices and the limitations of current device technology, such as low coherence times, quantum simulators provide a means for validating new and possibly deeper quantum algorithms.
There are various approaches to simulate a quantum circuit, the two main ones that provide exact results are: \textit{state vector} simulation \cite{fang2022efficient} and \textit{tensor network} contraction \cite{Gray2021}, which are detailed in the following subsections.

We aim at understanding how the inherent structure of a quantum circuit can affect the execution time, so as to preemptively identify which simulation strategy works best for which kind of circuit, by making use of \textit{ad hoc} metrics. These metrics provide a description of the overall structure of the quantum circuit, highlighting critical areas for the improvement of modern simulators. It will be possible to infer that any other quantum circuit, reflecting the characteristics provided in this paper, will scale similarly in terms of simulation.
The main current limitation of state vector simulators is the inherent exponential memory blowout linked to the system size, that has been tentatively compensated through state vector compression \cite{simulations_on_summit}, however the distributed application of vector-matrix multiplications still scales exponentially on the system size. Tensor networks have already shown promising results, with useful applications in the field of verification of real quantum computer's outputs, however due to the limited exploitation of the internal structures of the circuit-derived graph representation, the contraction path used is not granted to be optimal.
%\todo[inline]{The selection of the simulation approach is, currently, based on .... We aim at investigating the limits of each approach and to identify the more efficient simulation strategy off-line, by considering the quantum circuit characteristics.... o qualcosa del tipo.}
%\todo[inline]{I would have put details about the questions we are addressing, not just performance per se, but also what q circuits are meaningful, what are the limitations of current methodologies (TN vs. SV) and then how far we can go with these techniques not just memory or time but what we can actually simulate. }
\subsection{State vector simulation}
\begin{figure*}[!ht]
    \centering
    \includegraphics[height=0.22\linewidth]{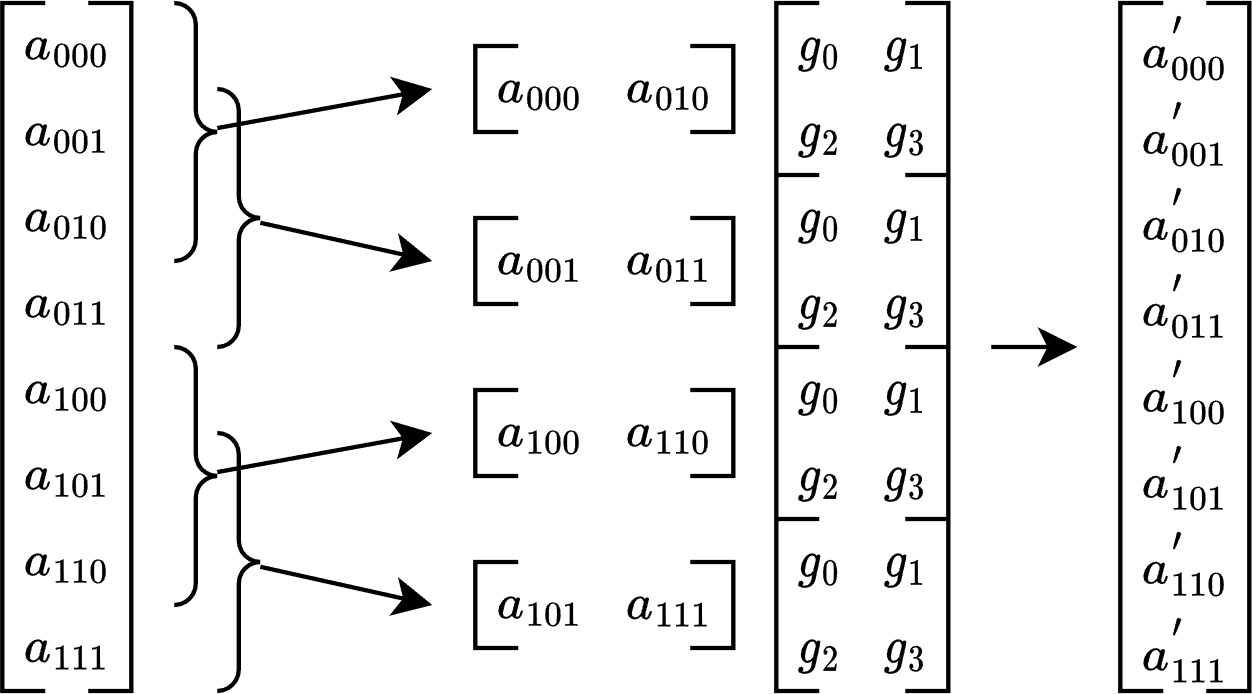}
    \quad\quad\quad\quad % FARQUAD
    \includegraphics[height=0.22\linewidth]{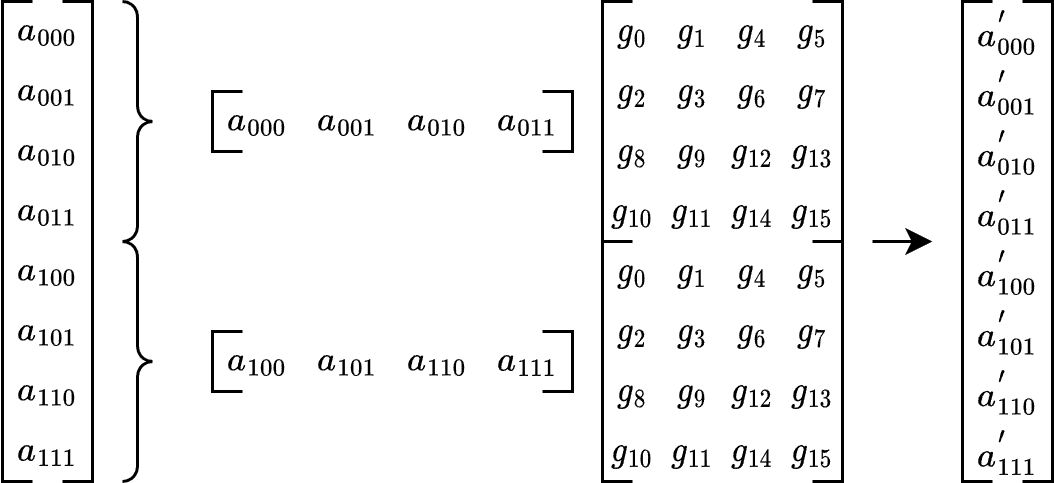}
    \caption{Splitting of the state vector and application of the vector-matrix multiplication in a 3 qubit system. On the left, a single qubit gate is applied to qubit 1, so the amplitude pairs are grouped by following a \textit{0-1} repeated scheme for the amplitude's index. On the right, a double qubit gate is applied on qubits 0 and 1, so the amplitude pairs are grouped following a \textit{00-01-10-11} scheme. All amplitude indices are written in little endian and ordered top to bottom.}
    \label{statevector_explained}
    
\end{figure*}

% How does this method work?
% Memory and matrix multiplication distribution bottlenecks
% Immediate availability of the whole wavefunction
Quantum states are represented by a wave function, which can be encoded into a state vector. Given any quantum system of $N$ qubits, its corresponding state vector will contain $2^N$ complex probability amplitudes, one for each possible output bit string. Quantum gates are applied by splitting the state vector into smaller vectors of size equal to that of the gate to be applied, then each sub-vector is multiplied with the gate matrix and the resulting sub-vectors are reassembled in the evolved state vector. The splitting operation is performed according to the qubits over which the operator is applied. This can be intuitively understood by considering the ordered set of output bit strings: the probability amplitudes corresponding to a given sequence of qubits, which depends on the qubit indices the operator acts onto, are grouped together. An example of this process for both single and two qubit gates is depicted in Figure \ref{statevector_explained}.

\begin{figure*}[!ht]
    \centering
    \includegraphics[height=.16\linewidth]{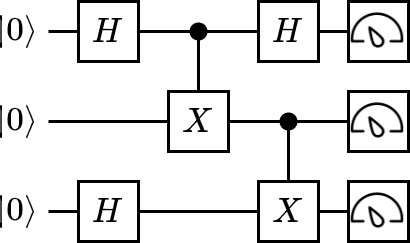}
    \quad\quad\quad\quad % FARQUAD
    \includegraphics[height=.16\linewidth]{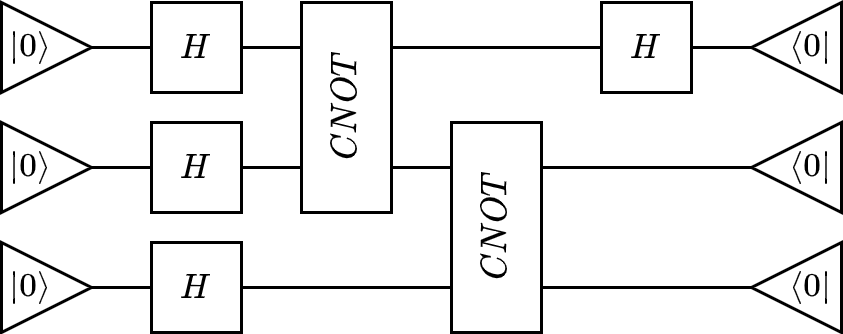}
    \quad\quad\quad\quad % FARQUAD
    \includegraphics[height=.16\linewidth]{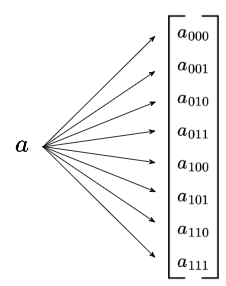}
    \caption{The process of converting a quantum circuit into a tensor network to extract a probability amplitude. On the left, an example of a quantum circuit. In the center, the circuit gets converted into a tensor network representation, where single and double qubit gates become order-2 and order-3 tensors, respectively. On the right, after the tensor network contraction, we get the probability amplitude of a specific bit string: by repeating this process over all output bit strings, one may reconstruct the whole state vector.}
    \label{tensor_network_explained}
    \vspace{-10pt}
\end{figure*}

The state vector simulation's complexity scales linearly in time with respect to the number of gates \cite{fang2022efficient}. However, the memory footprint of the state vector and the number of vector-matrix multiplications performed increase exponentially with the number of qubits present in the system to be simulated, so this approach is not scaleable indefinitely.
To put that into perspective, it is possible to roughly estimate  the number of atoms in the observable universe to be $10^{82} \approx 2^{270}$~\cite{plack_collaboration}: this means that, if we were to to store a single amplitude value inside each of them to represent a state vector, we could only represent state vectors of systems with up to 270 qubits. Well known quantum algorithms need significantly more logical qubits \cite{qubits_needed_for_shor}, and this is without considering the cost in terms of classical computation time, which may add up to reach unfathomable time scales\cite{Arute2019}.
Overall, the state vector approach is generally convenient when simulating small quantum systems, as it produces a full description of the output wavefunction.

\subsection{Tensor network simulation}
% While keeping the same circuit representation, the circuit's topology is read as that of a graph
% The resulting topology is encoded in that of a max-cut problem, where a candidate contraction path is computed
% The network is contracted by leaving the indices for the output bitstring open
% The final tensor's indices are then closed over an output bitstring and it gets contracted, resulting in the probability amplitude associated to that output bitstring
% The memory occupancy grows linearly with the number of gates present in the quantum circuit
% The pathfinding operation is an NP-hard problem, so there is no guarantee to find a good path, nor that the contraction takes less time than actual statevector simulation
% The full statevector is not available, but the final bitstring contraction must be repeated for each bitstring
Quantum gates and quantum basis states are represented by tensors. The graphical representation of a quantum circuit can be read as a directed acyclic graph (DAG), where the vertices are represented by quantum gates or basis states and the edges are represented by the qubit \textit{wires}. The input tensors are the basis states, encoded as follows
\begin{equation}
    \ket{0} = \begin{bmatrix}
        1 \\ 0
    \end{bmatrix}
    , \quad
    \ket{1} = \begin{bmatrix}
        0 \\ 1
    \end{bmatrix}
\end{equation}

All other gates use their standard matrix representation. The contraction of an edge corresponds to the multi-dimensional generalisation of the dot product between two tensors over a shared index. The measurement operators at the end of the quantum circuit are substituted by open indices. Whenever a full network contraction operation is performed, the output open indices are closed with the conjugate tensors of the basis states that encode a specific bit string. Doing so, followed by the contraction of the newly closed indices, produces the probability amplitude of the chosen bit string. Figure \ref{tensor_network_explained} provides graphical insight over this process.

It is possible to use tensor networks to reconstruct the whole state vector, by closing and contracting the output indices over different bit strings, however doing so incurs in the same limits of the state vector simulator for storing the final vector.

\obs{
    Tensor network contractions can reconstruct the full state vector, and such process can be trivially parallelised over different bitstrings.
}

The memory occupancy of the tensor network grows linearly with respect to the number of quantum gates and qubits in the system.
This approach moves the complexity of simulation to that of finding an optimal contraction path for the tensor network, which is known to be an NP-hard problem \cite{goemans1995}.
Efficient heuristics specialised for quantum circuit-derived tensor networks have been proposed, however there is no \textit{catch all} solution for this kind of problem.
The pathfinding algorithm used in this work \cite{Gray2021}, despite representing the state-of-the-art for this class of problems, only strives to optimise having the lowest possible amount of floating-point operations across the whole contraction process, which does not prevent the formation of large intermediate tensors, something that inevitably reduces contraction performance. Besides, as it will be discussed in Section \ref{pathfinding_contraction_performance}, the optimiser may easily be locked in a local minimum in some problems, whereas other problems feature smoother landscapes in terms of pathfinding complexity.
If the contraction path is not optimal, it may lead to increased computation time, possibly making it less efficient than state vector simulation altogether. To the interested reader, we suggest some resources for tensor network theory by J. Biamonte \cite{biamonte2017tensor, biamonte2020lectures}.

\section{Benchmarks and metrics}
\label{benchmarks_metrics_sec}
% The circuits used in this test are widely recognised and used in the quanutm computing field
% Some have been featured in the cuQuantum-benchmarks suite, others are acknowledged in the SupermarQ and QASMBench suites
% Table of benchmarks considered, area of usage and possible requirements
To assess the performance of current state of the art simulators and to select the most efficient one, it is necessary to use a set of metrics and quantum circuits which are relevant and well established in the quantum computing field. 
We rely on two quantum circuit benchmarking suites, which are widely recognised in the literature: SupermarQ \cite{supermarq} and QASMBench \cite{qasmbench}. Both of these suites provide their own sets of quantum circuits, that, however, have been specifically selected for testing the hardware performance of real quantum devices. For this reason, some of these circuits boast little to no practical use in the context of noiseless exact simulation, such as the error correcting code circuits in SupermarQ, or the Greenberger–Horne–Zeilinger.

\obs{
    Not all quantum circuits generally used for benchmarking are computationally representative in a classical simulation environment.
}

Furthermore, both suites introduce a list of metrics that characterise the topological nature of static quantum circuits. These metrics provide a measure of topological properties of the graph derived from the quantum circuit representation, letting us correlate such properties with the runtime performance statistics.
%\todo[inline]{explain why and how you use metrics for your study here.}

\subsection{SupermarQ}
% Short introduction of the metrics used, brought in from SupermarQ
In the SupermarQ \cite{supermarq} suite, six metrics are introduced, however we will only consider the ones that have topological significance, referring to all elements which may alter the circuit-derived graph, that is the relative presence and the disposition of two-qubit or higher size quantum gates. All metrics range in $[0, 1]$, where higher is closer to $1$.

\subsubsection{Program communication}
This metric measures the amount of interconnections present in a quantum program, computed as the ratio of the average degree of interaction of the quantum circuit in graph form with that of a maximally connected graph with a number of nodes equal to the number of qubits in the circuit. The term $d(q_i)$ is the degree of the \textit{i-th} qubit.
\begin{equation}
    PC = \frac{\sum_i^N d(q_i)}{N(N-1)}
\end{equation}
\subsubsection{Critical depth}
The critical depth represents the ratio between the longest chain of two-qubit operators and the total number of two qubit gates in the circuit. It gives a measure on whether the program's output heavily relies on distributed entanglement or not. $n_{e_d}$ is the total number of two qubit gates on the circuit's critical depth path, while $n_e$ is the number of two qubit gates in the circuit.
\begin{equation}
    CD = n_{e_d} / n_e
\end{equation}

\subsubsection{Entanglement ratio}
This measure is the ratio of the number of entanglement operators, $n_e$, with the total number of gates in the quantum circuit, $n_g$.
\begin{equation}
    E = n_e / n_g
\end{equation}

\subsubsection{Parallelism}
It is a measure of the number of concurrent operations made in the same time step, intuitively understood as the degree of \textit{compression} of the quantum circuit. The number of gates $n_g$ is compared with the depth $d$ of the program, then such value is normalised with respect to the number of qubits $n$.
\begin{equation}
    P = \left( \frac{n_g}{d} - 1 \right) \frac{1}{n - 1}
\end{equation}

\subsection{QASMBench}
% Short introduction of the metrics used, brought in from QASMBench
The metrics introduced in the QASMBench \cite{qasmbench} suite are more tied to the architectural implementation of physical quantum devices. Follows the definition of the only topologically significant metric.

\subsubsection{Entanglement variance}
This metric defines the spread of entanglement among the qubits in the quantum circuit. It checks whether there are a few qubits which feature most of connections towards the others, or if all qubits are sharing the same amount of entangling connections. In a quantum program with $N$ qubits, the number of two-qubit gates acting on the \textit{i-th} qubit is $n_{g_2}(q_i)$, while the average number of two-qubit gates per qubit is $\overline{n_{g_2}}$.
\begin{equation}
    EV = \frac{\log(\sum_{i=0}^{N} (n_{g_2}(q_i) - \overline{n_{g_2}} )^2 + 1) }{N}
\end{equation}

\begin{table*}[!t]
    \centering
    \caption{The quantum circuits used as benchmarks for the evaluations of this paper.}
    \label{circuits_list_table}
    \begin{tabular}{| m{60pt} | m{160pt} | m{120pt} | m{100pt} | m{20pt} |}
    \hline
        \vspace{2pt}\centering\textbf{Circuit name} & \vspace{2pt}\centering\textbf{Description} & \vspace{2pt}\centering\textbf{Total gates} & \vspace{2pt}\centering\textbf{Total multi-qubit gates} & \vspace{2pt}\textbf{Ref} \\[1.1ex] \hline\hline
        QAOA & Quantum Approximate Optimisation Algorithm & \vspace{2pt}$\frac{3}{2}PN(N-1)+2N$ & \vspace{2pt}$PN(N-1)$ & \cite{qaoa_ref} \\[1.1ex] \hline
        Random & Google quantum supremacy circuit & \vspace{2pt}$(1-k)(N(\lfloor N/2 \rfloor + N\%2)) + kN^2$ & \vspace{2pt}$kN(\lfloor N/2 \rfloor)$ & \cite{Arute2019} \\[1.1ex] \hline
        QPE & Quantum Phase estimation & \vspace{2pt}$\frac{N(N-1)}{2} + 2N - 1 + \lfloor \frac{(N-1)}{2} \rfloor$ & \vspace{2pt}$\frac{(N^2 - N)}{2} + N - 2 + \lfloor \frac{(N-1)}{2}  \rfloor$ & \cite{qpe_ref} \\[1.1ex] \hline
        QFT & Quantum Fourier transform & \vspace{2pt}$\frac{1}{2}N(N+1) + \lfloor N/2 \rfloor$ & \vspace{2pt}$\frac{1}{2}(N^2-N) + \lfloor N/2 \rfloor$ & \cite{qft_ref} \\[1.1ex] \hline
        VQE & Variational Quantum Eigensolver & \vspace{2pt}$L(5N-1)+N$ & \vspace{2pt}$L(N-1)$ & \cite{vqe_ref} \\[1.1ex] \hline
        Hamiltonian sim. & One-dimensional Hamiltonian time evolution & \vspace{2pt}$3T(2N-1)$ & \vspace{2pt}$T(N-1)$ & \cite{supermarq} \\[1.1ex] \hline
        Hidden Shift & Find the shift $s$ such that $g(x)=f(x+s)$ & \vspace{2pt}$3N + 2M + \lfloor N/2 \rfloor $ & \vspace{2pt}$\lfloor N/2 \rfloor $ & \cite{hidden_shift_ref} \\[1.1ex] \hline
        Bernstein-Vazirani & Hidden bit string extraction & \vspace{2pt}$2N+M$ & \vspace{2pt}$M$ & \cite{bernstein_vazirani_ref} \\[1.1ex] \hline
    \end{tabular}
    
\end{table*}

\begin{figure}
    \centering
    \includegraphics[width=\linewidth]{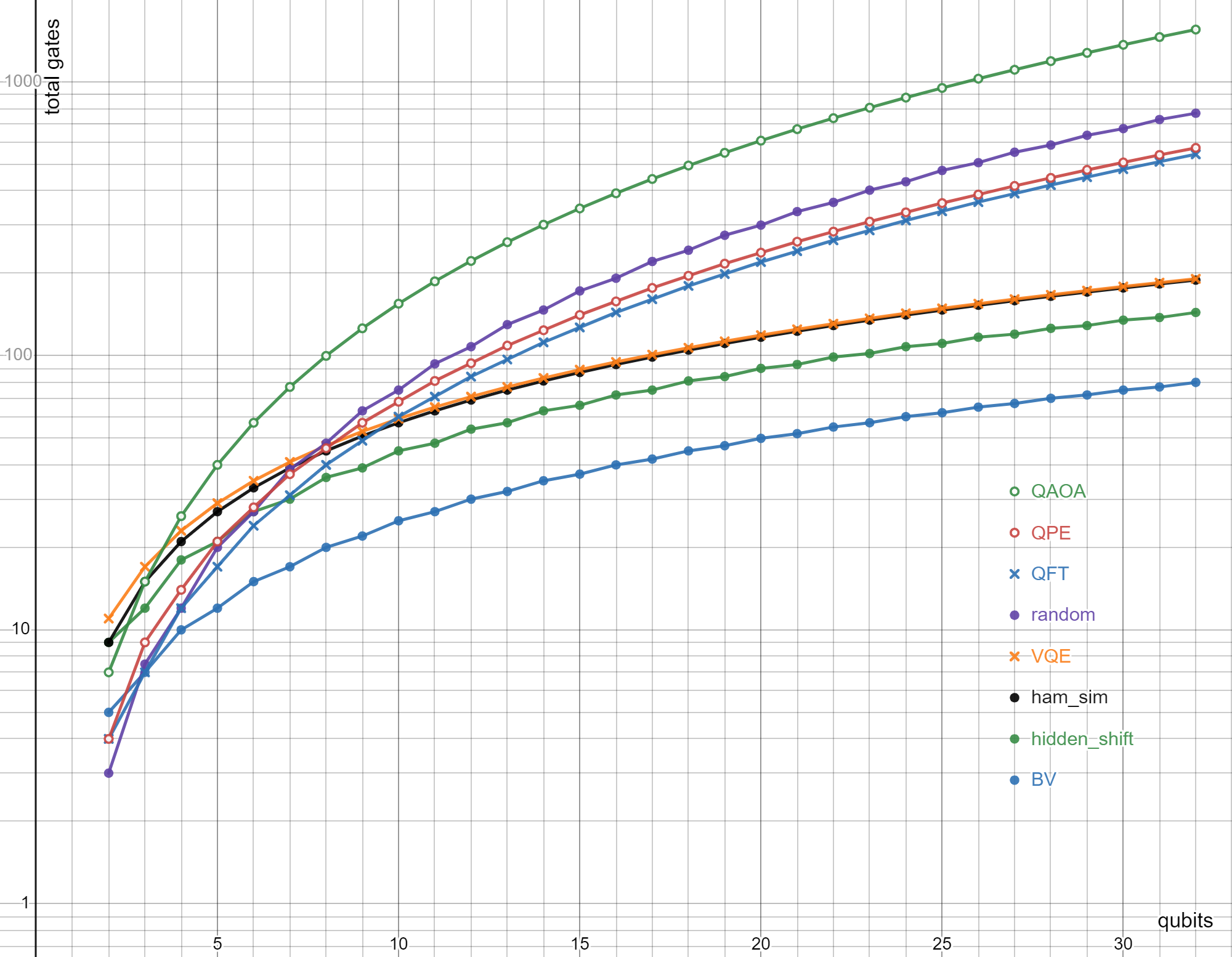}
    \caption{The scaling of the number of total gates of the circuits considered in this paper, from size 2 to size 32. The Y-axis is in logarithmic scale. The controlling variables are set as $P=1$, $k=0.5$, $M=\lfloor kN \rfloor$, $L=1$, $T=1$.}
    \label{scaling_number_gates}
    
\end{figure}

\subsection{Benchmark circuits}
% \hl{In order to provide an extensive evaluation we consider a broad set of circuits with different characteristics...}
In order to provide a broad, extensive and scalable evaluation, we consider a specific set of circuits,
selected to encompass some of the applications for quantum computing that do not leverage the presence of quantum noise. As such, their results are significant in terms of exact theoretical simulation. The list of circuits, with information regarding usage, scaling of the number of gates and references, is detailed in Table \ref{circuits_list_table}. All circuits considered can be freely expanded over any problem size, making them easily adaptable to benchmark future hardware and simulation platforms. For the sake of the hereby presented analysis, we tested all circuit qubit sizes in the range $[2-32]$.

\subsubsection{QAOA}
The \textit{Quantum Approximate Optimisation Algorithm} is a variational circuit that use all-to-all connectivity to encode classical problems, such as the max-cut problem. The algorithm version used in this paper is the vanilla one \cite{qaoa_ref} with parameter $P=1$, although other versions exist, such as the one with ZZ-Swap gates \cite{qaoa_swap_network}.

\subsubsection{Random}
The \textit{Random} quantum circuit, notably dubbed \textit{Quantum Supremacy circuit} by the GoogleAI group that introduced it \cite{Arute2019}, is composed of multiple repeated layers of random gates picked from the set $\mathcal{G} = \{H, X, RZ, RX, RY, CNOT, CZ, SWAP\}$. The number of layers has been set equal to the number of qubits in the system. Given the random nature of the circuit, it is possible to define a lower and an upper bound for the number of gates that can be found in the circuit.
This circuit has been purposefully built to avoid any internal structure, so as to be as complex as possible to simulate. Despite achieving such a goal, the applicability for this subroutine to real world problems remains questionable at best.

\subsubsection{QPE}
The \textit{Quantum Phase Estimation} subroutine is one of the foundational steps in Shor's algorithm \cite{qpe_ref}\cite{shor1997}, able to solve the order finding problem of a modulo function.

\subsubsection{QFT}
The\textit{ Quantum Fourier Transform} \cite{qft_ref} is one of the most widely known quantum subroutines, which uses phase encoding to efficiently perform the Fourier transform. This quantum circuit is a good candidate for simulation, since it is built by the recursive application of the same operator, possibly giving rise to exploitable internal structures.

\subsubsection{VQE}
The \textit{Variational Quantum Eigensolver} is an hybrid quantum-classical algorithm that uses iterative optimisation to find the ground state of a molecule encoded in a quantum register. The circuit is built by repeating \textit{L} times a given structure. For the sake of simplicity, the version used in this paper assumes $L=1$.
Its applications focus mainly on, but are not limited to, the simulation of the bond energies of chemical compounds.

\subsubsection{Hamiltonian Simulation}
This circuit encompasses a general approach for the encoding and simulation of the time evolution of a given Hamiltonian in a quantum computer. The circuit is characterised by the repeated application of a quantum subroutine over $T= \textit{total\_time} / \textit{time\_step}$ iterations.
The benchmark we consider computes the magnetic interactions  of a monodimensional chain of spins.

\subsubsection{Hidden Shift}
The \textit{Hidden Shift} quantum circuit is able to find the value $s$ of a function $g(x) = f(x + s)$ by performing a single query to the function, leveraging the superposition of all the possible inputs. The term $k$ represents the percentage of bits equal to 1 in the binary representation of the shift value.

\subsubsection{Bernstein-Vazirani}
This quantum algorithm can solve the problem of finding out the bit string that satisfies a given function by performing a single query, whilst the classic algorithm would require at most $N$ queries, where $N$ is the total number of possible bit strings. Once again, the term $k$ is the percentage of bits set to 1 in the binary representation of the solution.

\section{Results}
\label{results_sec}
In order to find an answer to the research questions defined in Section \ref{intro}, we will start by characterising the quantum circuits according to the metrics introduced in Section \ref{benchmarks_metrics_sec}, then we will perform various simulations, with the objective to relate the metrics with execution time, memory occupancy, and in the specific case of tensor networks, distributed sliced contraction performance and pathfinding efficacy.
The library used for running the simulations is the NVIDIA cuQuantum library (version v24.03)\cite{cuquantum}, adapted to the specific needs of our analysis.
This gives us access to three different GPU accelerated simulation backends:
\begin{itemize}
    \item \textit{qsim-cusv}: a state vector simulator that uses the \textit{cuStateVec} backend.
    \item \textit{qsim-cuda}: a state vector simulator that uses the \textit{cupy} backend.
    \item \textit{cutn}: a tensor network simulator that uses the \textit{cuTensorNet} backend for contraction.
\end{itemize}
All experiments have been run on CINECA's Leonardo supercomputer. Apart from the distributed experiments, all other experiments have been run on a single node, using 8 cores of an Intel Xeon Platinum 8358 CPU, 128 GB of RAM and one NVIDIA Ampere A100 64 GB GPU.
Simulations have all been limited to a problem size of 32 qubits, as that is the largest statevector that can be represented in a single available GPU, although larger tensor networks could indeed be simulated. 
Given the high computational cost of state vector and tensor network methods, CPU-based simulation algorithms will not be considered for this analysis, since it would not provide any meaningful comparison in terms of performance.
\begin{figure*}[!hpt]
\begin{minipage}{0.49\textwidth}
  \centering
  \includegraphics[height=0.23\textheight]{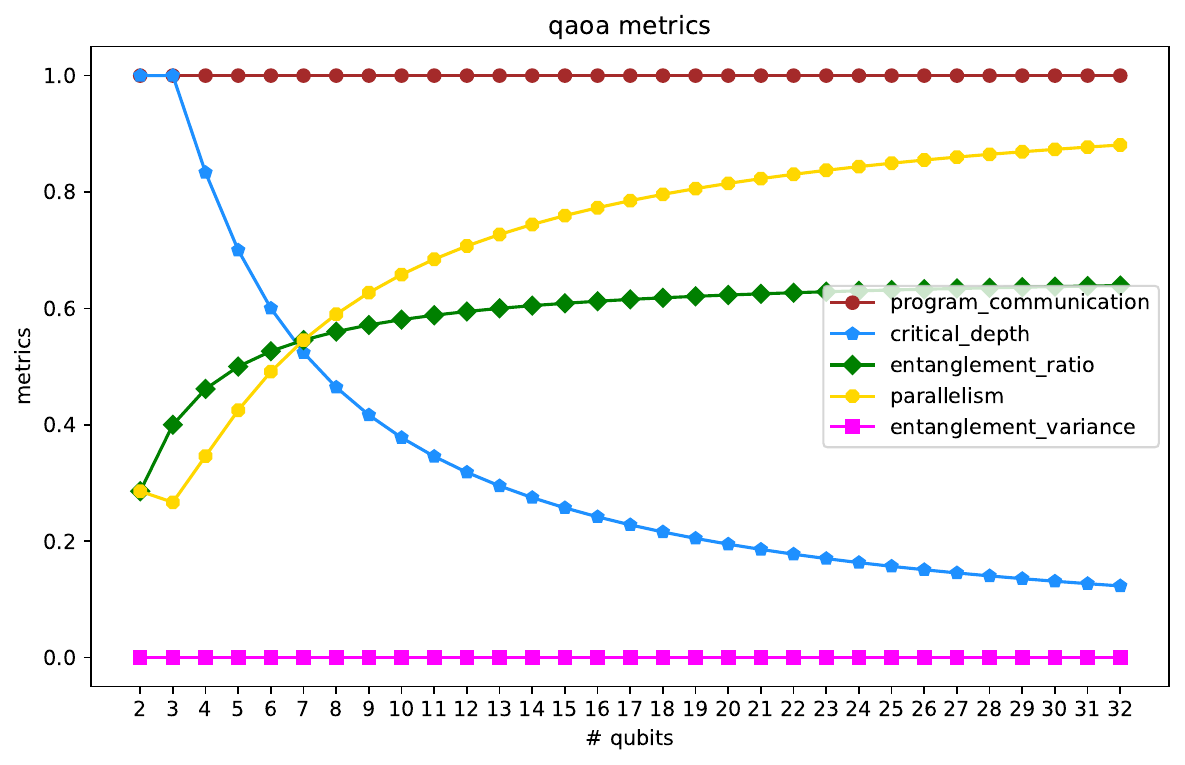}
  % \caption{QAOA circuit metrics}
  \label{qaoa_metrics}
\end{minipage}%
\begin{minipage}{0.49\textwidth}
  \centering
  \includegraphics[height=0.23\textheight]{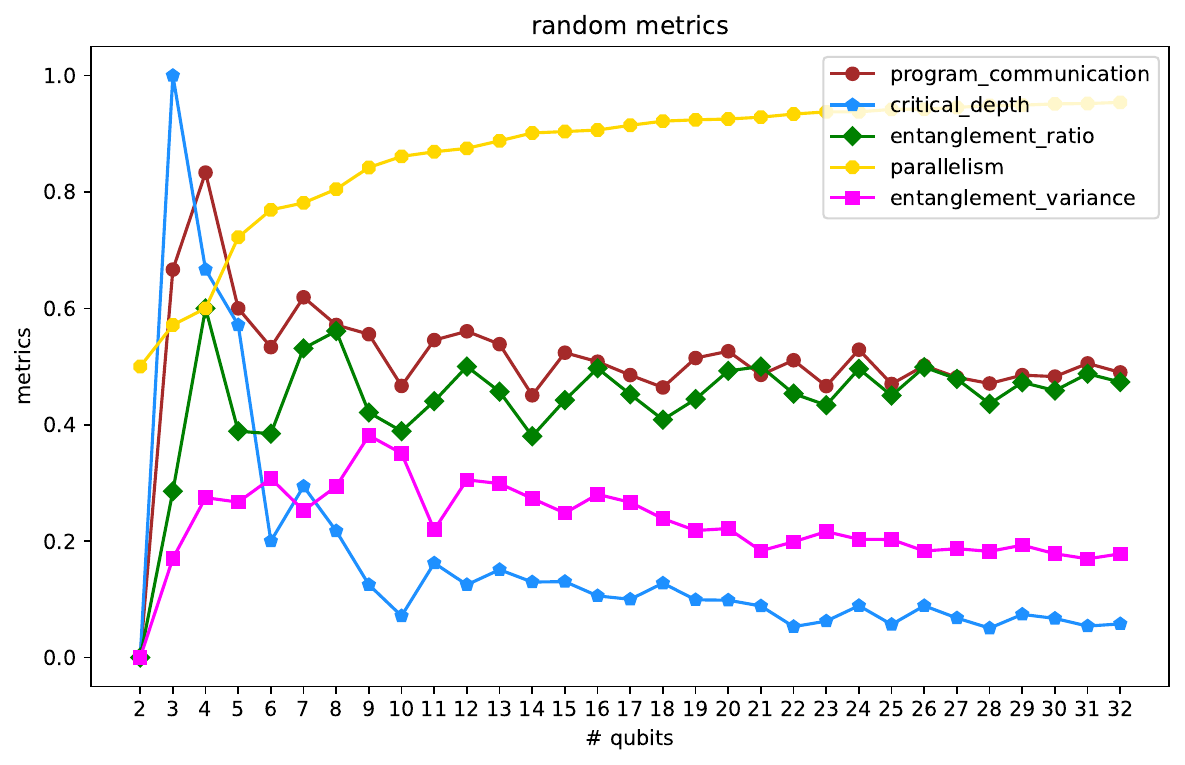}
  % \caption{Random circuit metrics}
  \label{random_metrics}
\end{minipage}
\begin{minipage}{0.49\textwidth}
  \centering
  \includegraphics[height=0.23\textheight]{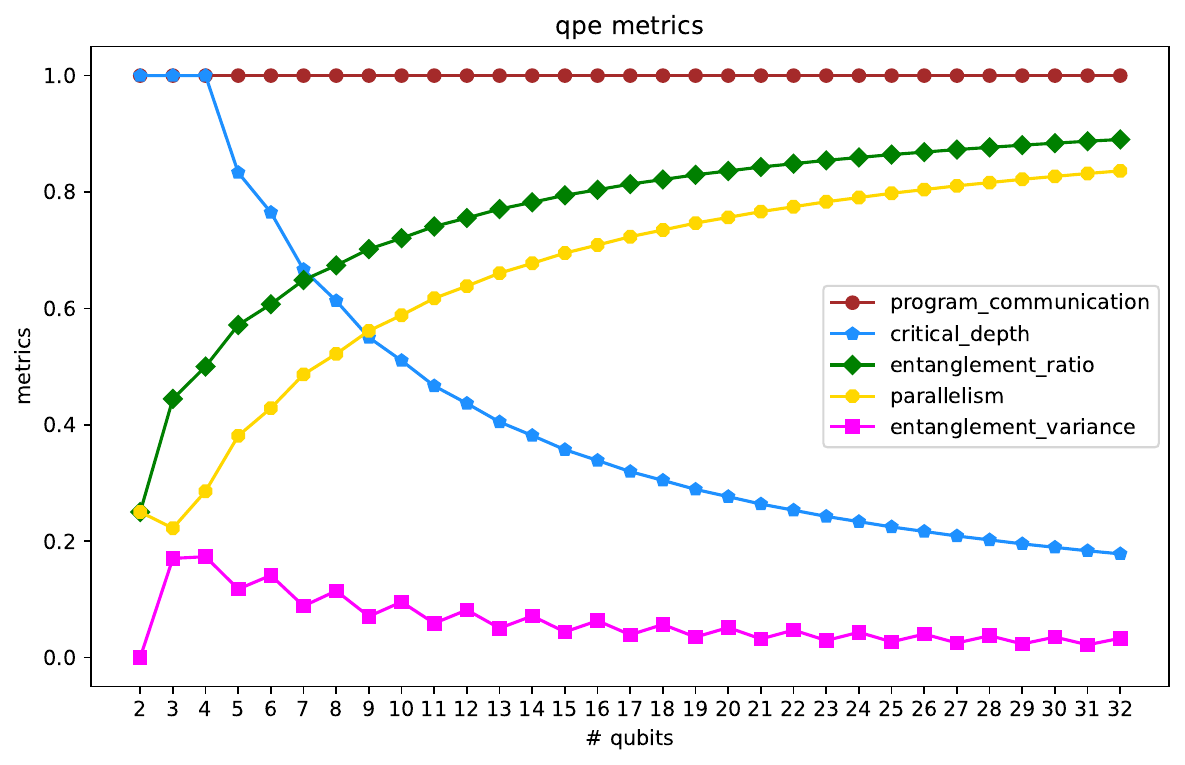}
  % \caption{QPE circuit metrics}
  \label{qpe_metrics}
\end{minipage}%
\begin{minipage}{0.49\textwidth}
  \centering
  \includegraphics[height=0.23\textheight]{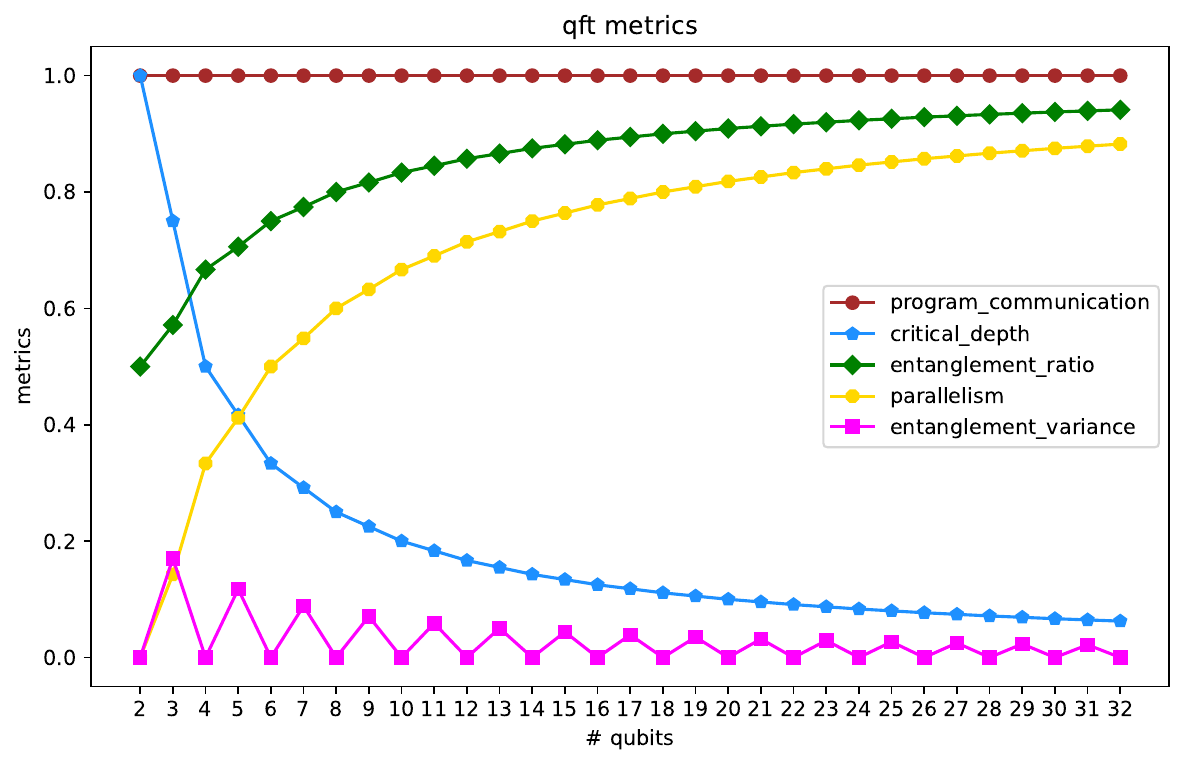}
  % \caption{QFT circuit metrics}
  \label{qft_metrics}
\end{minipage}
\begin{minipage}{0.49\textwidth}
  \centering
  \includegraphics[height=0.23\textheight]{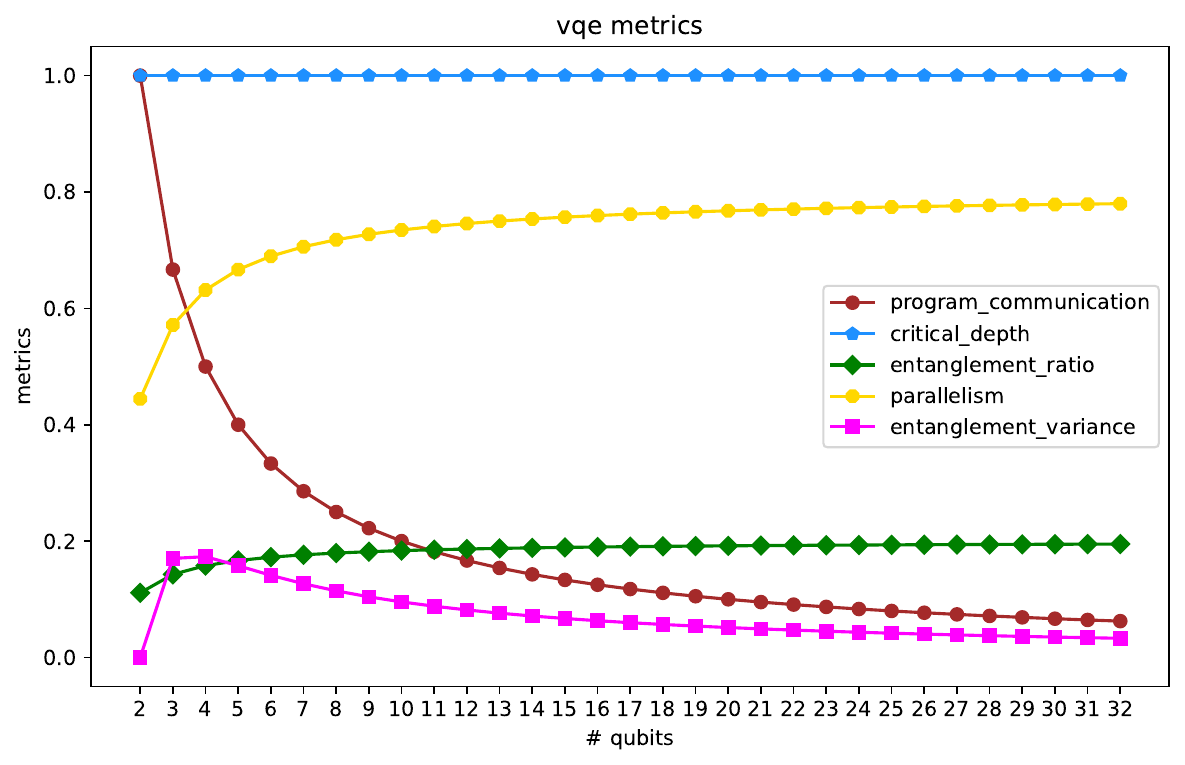}
  % \caption{VQE circuit metrics}
  \label{vqe_metrics}
\end{minipage}%
\begin{minipage}{0.49\textwidth}
  \centering
  \includegraphics[height=0.23\textheight]{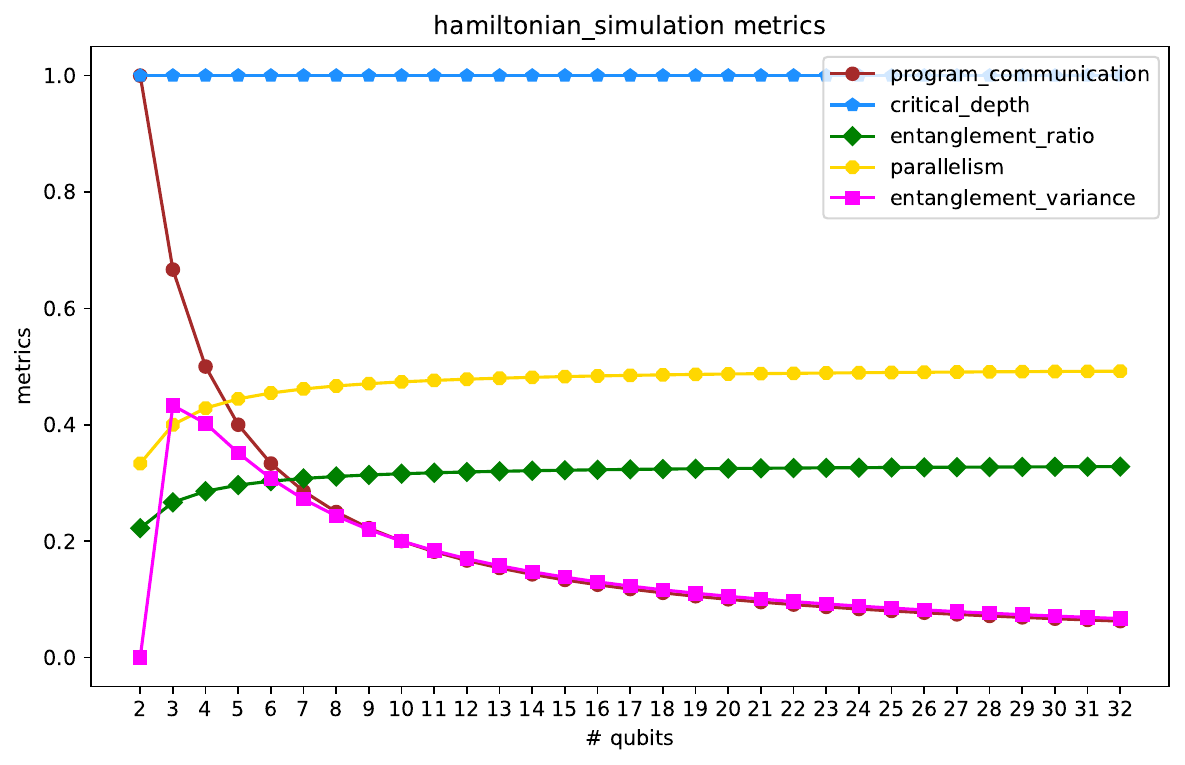}
  % \caption{Hamiltonian simulation circuit metrics}
  \label{hamiltonian_simulation_metrics}
\end{minipage}
\begin{minipage}{0.49\textwidth}
  \centering
  \includegraphics[height=0.23\textheight]{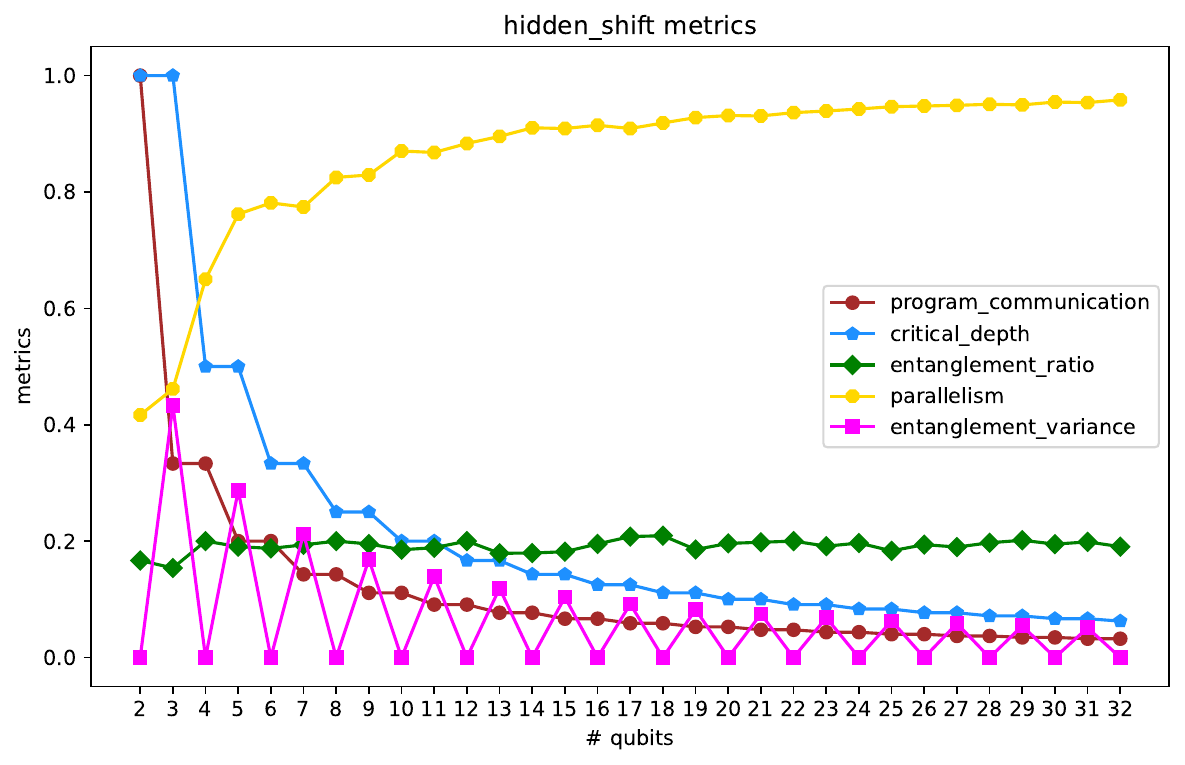}
  % \caption{Hidden Shift metrics}
  \label{hidden_shift_metrics}
\end{minipage}%
\begin{minipage}{0.49\textwidth}
  \centering
  \includegraphics[height=0.23\textheight]{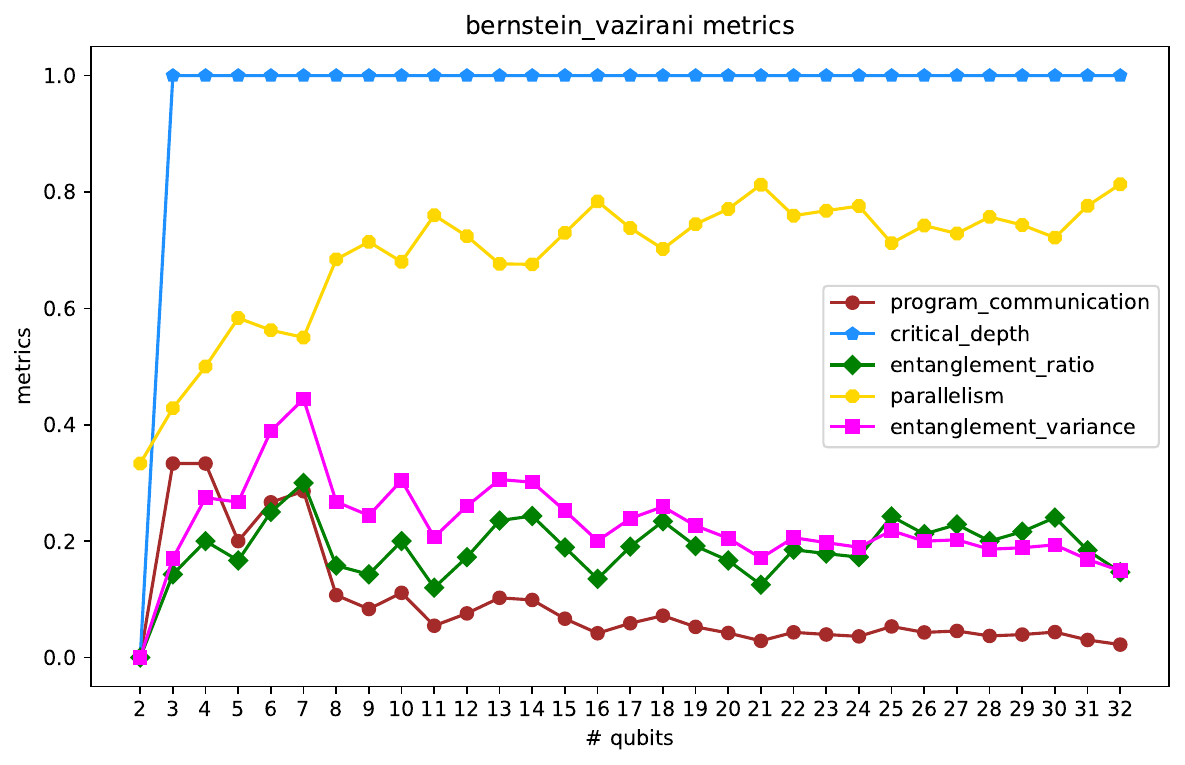}
  % \caption{Bernstein-Vazirani circuit metrics}
  \label{bernstein_vazirani_metrics}
\end{minipage}
\\
\caption{Metrics computed for all the benchmark circuits. Given that the topological structure of the Random, Hidden Shift and Bernstein-Vazirani circuits depends on an initialisation seed, the metrics for this circuit have been averaged over 100 different problem instances.}
\label{metrics_overall_image}
\end{figure*}

\begin{figure*}[!htb]
    \begin{minipage}{0.49\textwidth}
      \centering
      \includegraphics[width=\linewidth]{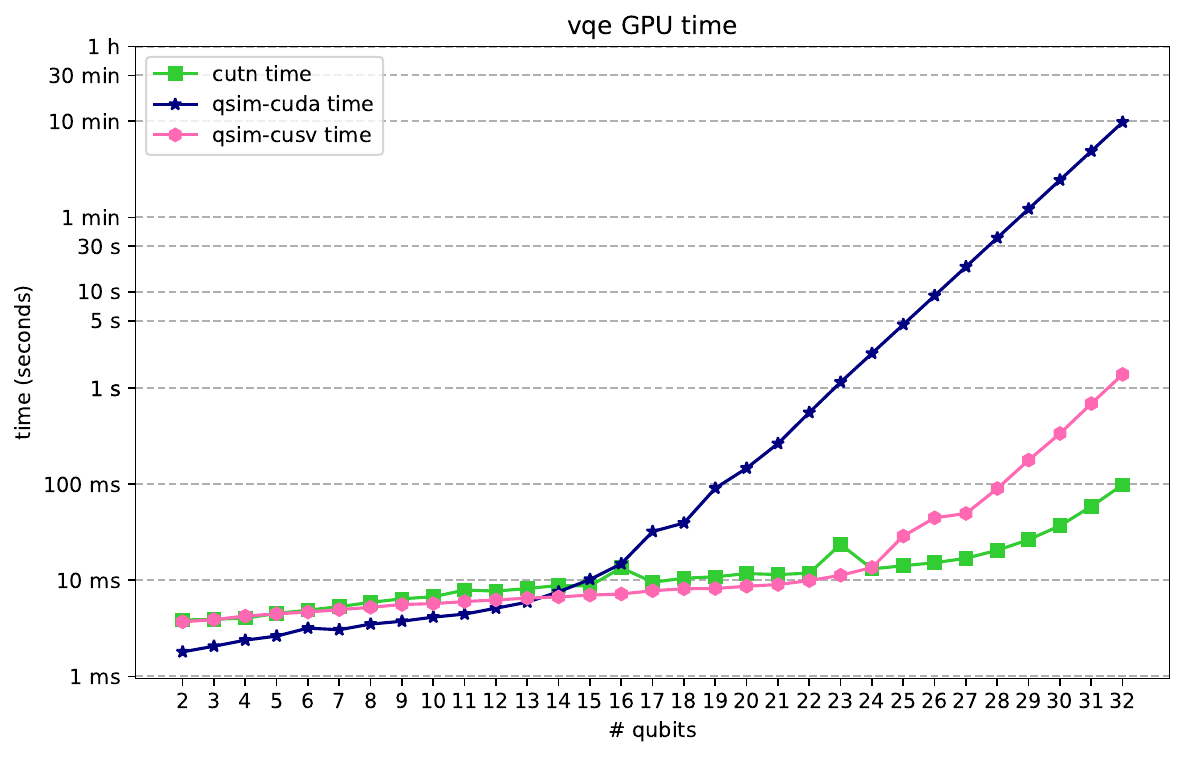}
      \label{vqe_performance}
    \end{minipage}%
    \begin{minipage}{0.49\textwidth}
      \centering
      \includegraphics[width=\linewidth]{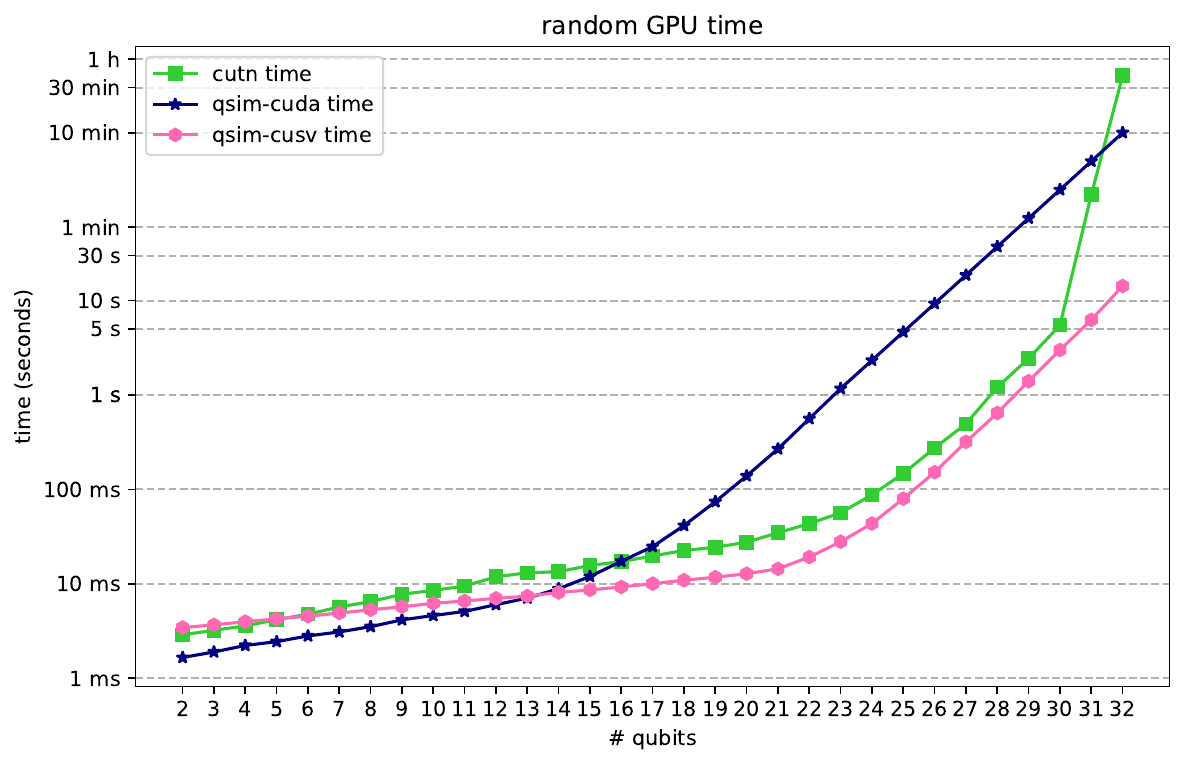}
      \label{random_performance}
    \end{minipage}
    % \begin{minipage}{0.46\textwidth}
    %   \centering
    %   \includegraphics[width=0.9\linewidth]{Figures/vqe_topology_32_qubits.png}
    %   \label{vqe_topology}
    % \end{minipage}%
    % \quad\quad\quad\quad
    % \begin{minipage}{0.46\textwidth}
    %   \centering
    %   \includegraphics[width=0.9\linewidth]{Figures/random_topology_32_qubits.png}
    %   \label{random_topology}
    % \end{minipage}
    \caption{
    % At the top, t
    The execution times for the VQE circuit (left) and the Random circuit (right). All data has been collected by doing 3 warmup runs and then averaging the results of 10 additional runs. The tensor network time is the sum of the pathfinding time, done in CPU, and the contraction time, done in GPU.
    % At the bottom, the network structure derived from the same circuits of size 32 qubits, elided of the tensors with degree $d=2$. Blue nodes represent input tensors, while red nodes represent output sensors, node size represents the order of the tensor.
    }
    \label{performance_topology_overall_image}
    \vspace{-10pt}
\end{figure*}

\begin{figure}[t!]
    \centering
    \includegraphics[width=\linewidth]{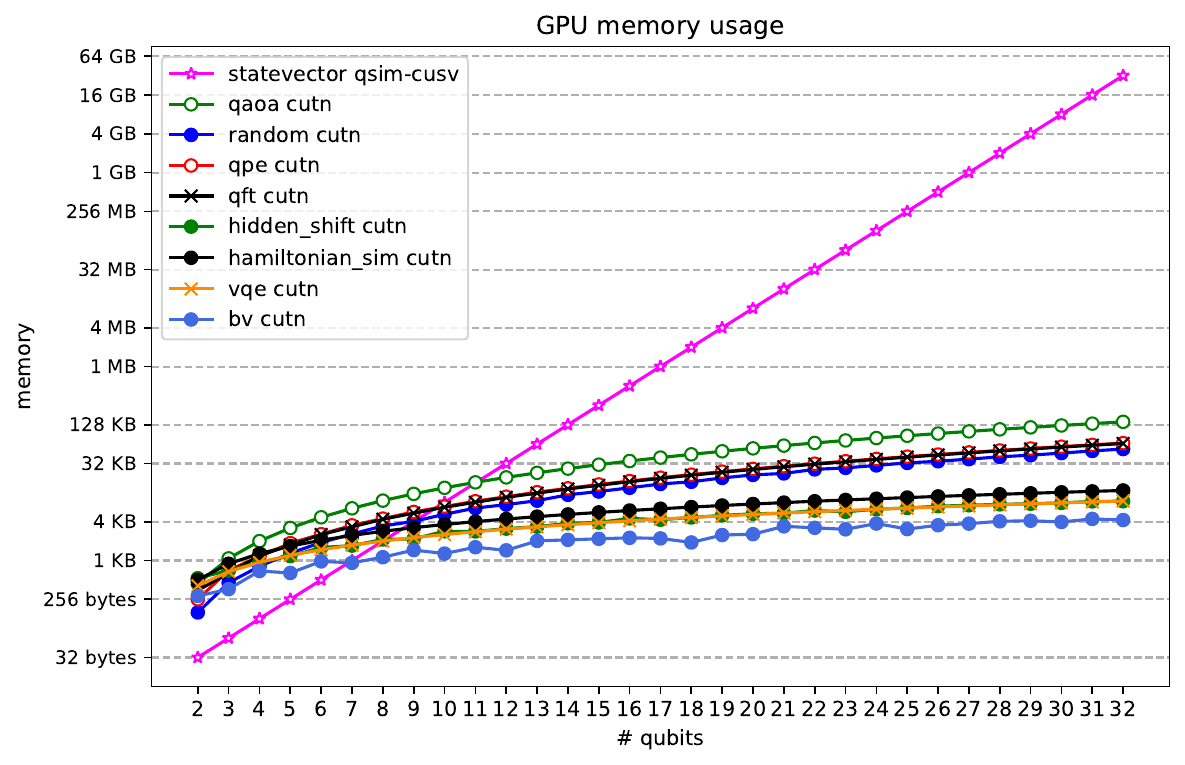}
    \caption{Memory occupancy of a general state vector and the circuit derived tensor networks.}
    \label{memory_occupancy_sv_vs_tn}
\end{figure}

\subsection{Quantum circuit properties}
\label{quantumCircuitProperties}
% List and comment the SupermarQ and QASMBench properties of the considered circuits     
Following the results reported in Figure \ref{metrics_overall_image}, we will analyse each metric independently. The metrics for the Random, Bernstein-Vazirani and Hidden Shift benchmarks have been averaged over 100 circuit samples, to compensate for the fact that these circuits do not have a constant topology.

% PROGRAM COMMUNICATION RESULTS
The \textit{program communication} keeps a constant value of $1$ for the QAOA, QPE and QFT circuits, suggesting that those three algorithms feature at least a two-qubit operation with each of the other qubits in the quantum register. This means that the resulting topology of the circuit will be that of a fully connected graph. All other circuits, on the other hand, quickly drop towards values proximal to $0.1$ as soon as the number of qubits in the system increases, meaning that most of the qubits do not interact directly. This can be explained by circuit structures where a small number of qubits interact with all of the others, or by circuit structures where all qubits interact only with their closest neighbours. The main outlier is the Random circuit, which stabilises at a value of about $0.5$, meaning that, on average, each qubit interacts with at least half of the total number of qubits in the circuit.

% CRITICAL DEPTH RESULTS
The \textit{critical depth} starts at value $1$ for most circuits at low qubit sizes and rapidly drops towards the range $[0.08,0.22]$ for the QAOA, QPE, QFT, Random and Hidden Shift benchmarks. This, together with the \textit{program communication} score, means that the highly entangled structure of the first three circuits is not due to a chain of two qubit operators. The Hidden shift and the Random circuits, having both low \textit{program communication} and \textit{critical depth} scores, imply that the derived graph structure is sparsely connected, with a few "central" qubits sporting most of the two qubit gates towards all other qubits. The remainder of the quantum circuits in the test suite maintain a constant value at $1.0$. This, together with the \textit{program communication} metric implies that the graph structure of the VQE, Hamiltonian simulation and Bernstein-Vazirani circuits can be reduced to that of a single chain of nodes.

% ENTANGLEMENT RATIO RESULTS
The \textit{entanglement ratio} attains its maximum value in the QPE and QFT benchmarks, as those circuits are mainly composed of two-qubit gates. The QAOA and Random circuits are composed from $40\%$ to $60\%$ by multiple qubit gates, with the former saturating at $60\%$ as the size of the system increases to 32 qubits, whilst the latter, given the its non deterministic structure, boasts an average of about $50\%$. The other circuits, the Hamiltonian simulation, VQE, Hidden shift and Bernstein-Vazirani, are mainly made of single qubit gates, which can get easily processed during tensor network contraction \cite{Gray2021}, as such their \textit{ER} scores are lower, ranging $[0.15-0.35]$.

% PARALLELISM RESULTS
The \textit{parallelism} metric grows with respect to the circuit size for all the algorithms considered, with initial values ranging from $0.0$ for the QFT to $0.5$ for the Random. Generally, however, the metric's value saturates to different levels, with the QAOA, Random, QPE, QFT and Hidden Shift circuits passing the threshold $P>0.8$ for systems sizes of 32 qubits, suggesting that the density of their derived topology is very high. On the other hand, the VQE and Bernstein-Vazirani circuits saturate in the range $[0.6-0.8]$, suggesting a slightly more sparse topology. The Hamiltonian simulation circuit saturates at value $P\approx0.5$, highlighting its dependence on sequential processing of quantum information and a lower topological density.

% ENTANGLEMENT VARIANCE RESULTS
The \textit{entanglement variance} rapidly approaches \textit{zero} for almost all circuits considered in the benchmark suite. Notably, the QAOA circuit has a \textit{constant} variance value of $0.0$, meaning that independently from the circuit size, the number of two qubit operators is evenly split amongst all the qubit in the system. The QPE, QFT, VQE and Hidden Shift algorthms see a rapid decrese in the metric's value, approaching the range $[0.0-0.05]$ for circuits of 32 qubits, again hinting at the fact that most of the qubits take part in a similar amount of multi-qubit operations. The only two exceptions are the Random and the Bernstein-Vazirani circuits, which instead have a higher value of $ER\in[0.18-0.25]$. In the case of the Random circuit, this is due to the fact that the structire of the circuit does not follow a predefined scheme, whilst in the Bernstein-Vazirani circuit it is directly depended on the number of 1s present in the solution binary bitstring of the oracle function.

\subsection{Single GPU simulation performance}
\label{simulationPerformance}
% Put the time and memory occupancy graphs, then comment on them
% Multiple time graphs (one for each interesting circuit)
% A single memory occupancy graph (with multiple TN-size graphs for each circuit), the statevector memory occupancy is the same for every quantum circuit
% Highlight where the Statevector performs better (small qubit sizes)
% Comment on the relationship between the hardware block size and the sudden computation time increase in the statevector simulators
% Highlight the tensor network speedup over large qubit circuits, then describe the inherent advantages of using tensor networks

Figure \ref{memory_occupancy_sv_vs_tn} details the memory requirements scaling for a general state vector simulator and the tensor network representations of all the circuits considered in our benchmark.

The memory occupancy for both the \textit{qsim-cuda} and the \textit{qsim-cusv} simulators is the same, as they both have to store in memory the whole state vector of complex probability amplitudes. Each probability amplitude is stored as a \textit{complex-64} single precision binary number, scaling exponentially in memory, since the size of the state vector is $N=2^n$, with $n$ being the number of qubits in the system. The state vector is updated when applying new quantum gates, however its size remains unaltered, regardless of how many subsequent operations are applied to it.

The tensor network representation instead stores the initial state as a sequence $(1,2)$ tensors, and each quantum gate as either a $(2,2)$ order-2 tensor, in the case of single qubit operators, or as a $(2,2,2,2)$ order-4 tensor, in the case of a controlled gate, adding two dimensions of size 2 for each additional input of the operator. As such, the size of the tensor network representation scales linearly with the number of gates in the quantum circuit and the number of qubits in the system.

In Figure \ref{performance_topology_overall_image}, we can see a side by side comparison of the execution times of the VQE and the Random circuits.
% , together with the tensor network structure derived from the circuit, after the removal of all order-2 tensors, for readability's sake. 
The time performance data has been collected, for each circuit configuration, as the average time over 10 runs after having performed 3 warmup runs. Thus we collectively performed $3244$ quantum circuit simulations for this experiment. 
Code and execution time data relative to these other circuits is available at \cite{paperRepo}.
We notice how the \textit{qsim-cuda} simulator has the overall fastest performance for circuits with size of 13 qubits or less, after which the performance of the simulator degrades significantly. This is mainly due to the fact that the shared memory size of the NVIDIA A100, the device we used for simulations, is only configurable up to 164 KB: the state vector of a 13 qubits system, at \textit{complex-64} single precision, occupies 64 KB. The performance for circuits of size equal to 14 still holds up, as the state vector needs 128 KB to be stored. However, as soon as the size increases, the L1 cache is not large enough to fit the whole state vector and the simulation gets hindered by memory transfers. From that point onward, the execution time for the \textit{qsim-cuda} scales exponentially with the system size.
Quantum circuits with more than 14 qubits perform better on either of the other two simulators. Notably, the \textit{qsim-cusv} simulator starts to lose performance for system sizes larger than 22 qubits. This is once again due to the properties of the A100 GPU, which has an L2 cache size of 40 MB, whilst the state vector size of a 22 qubit system is 32 MB. As soon as we increase the system size, the memory requirement doubles and exceeds the cache, forcing additional memory transfers that make the time performance once again scale exponentially in time.

\begin{figure*}[!htb]
    \centering
    \includegraphics[width=\textwidth]{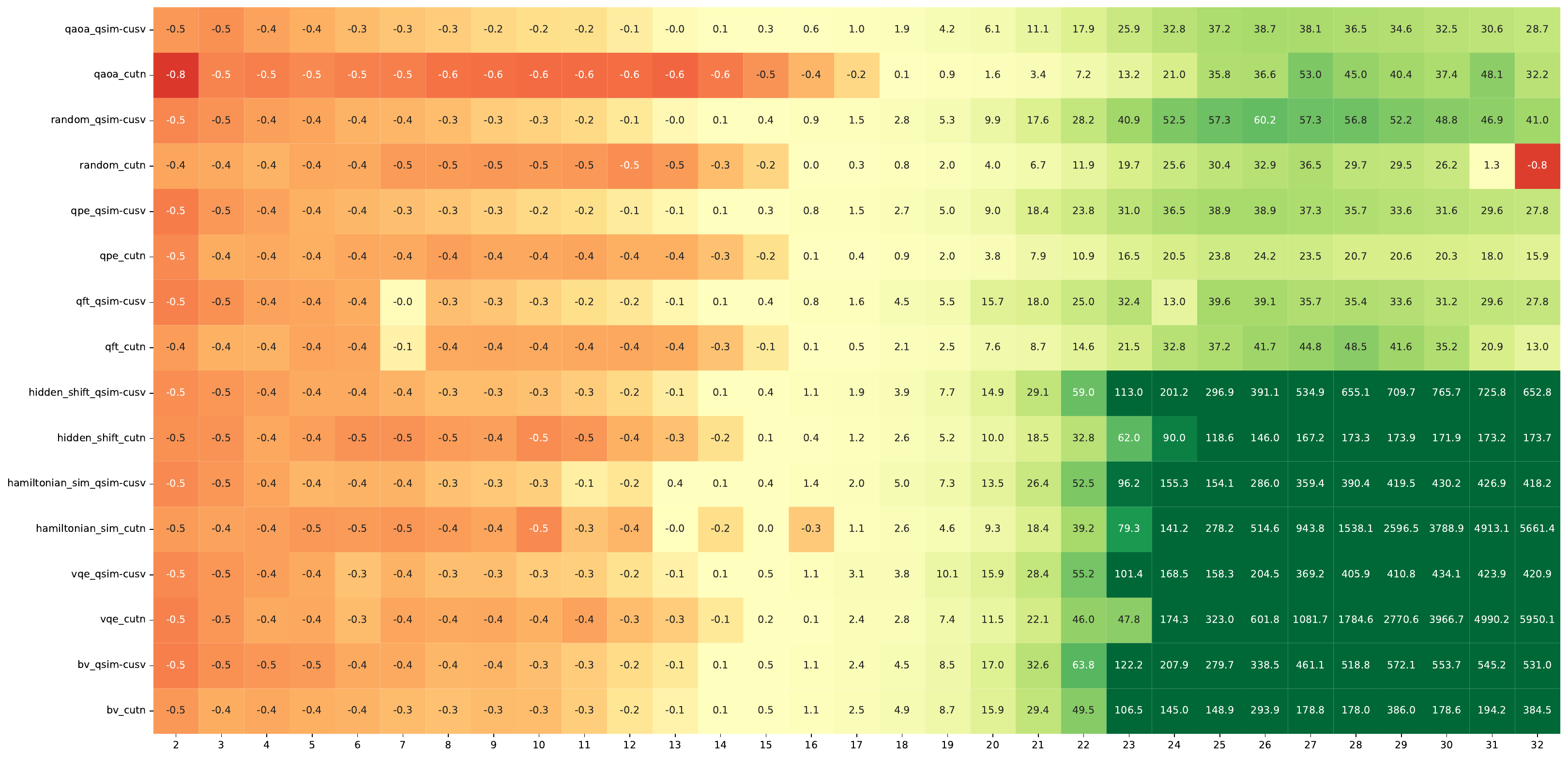}
    \caption{Heatmap of the speedup of the \textit{qsim-cuda} and the \textit{cutn} simulators with respect to the \textit{qsim-cuda} simulator.}
    \label{absolute_speedup_heatmap}
\end{figure*}

In the case of the Hamiltonian simulation, VQE and QAOA circuits, the tensor network outperforms both state vector simulators. This is a consequence of the fact that these circuits are mainly composed of single qubit gates, corresponding to an \textit{entanglement ratio} score lower than $0.5$, and have a well distributed amount of entanglement across the system, meaning an \textit{entanglement variance} score lower than $0.2$.
This leads to having small intermediate tensors during the contraction step, that eventually get merged together into larger tensors right towards the end of the process. As soon as one of these two metrics raises up too much, the pathfinding algorithm we employed \cite{Gray2021} struggles find an optimal contraction path.
Moreover, once all the order-2 tensors have been contracted, the resulting topology resembles that of a matrix product state (MPS), which is optimal for contraction. This is an important information, as the contractions of order-2 tensors are easy to perform and do not increase the intermediate tensor order during the contraction process.

\obs{
    Unstructured and unbalanced tensor networks give rise to large intermediate tensors during contraction in function of their qubit size and \textit{program communication} metric, hindering performance.
}

The circuit with the worst performance for the \textit{cutn} simulator is the Random circuit, which is the only problem where performance degrades by more than an order of magnitude, mainly due to its lack of an internal structure, something which can be clearly noticed on the right of Figure \ref{performance_topology_overall_image}. In all other benchmark circuits, the performance in terms of time is comparable to that of a state vector simulator, whilst keeping the reduced memory footprint of the tensor network representation. 

In Figure \ref{absolute_speedup_heatmap}, we can see the relative speedup of the \textit{qsim-cusv} and the \textit{cutn} simulators when compared to the \textit{qsim-cuda} simulator, whose baseline performance is mostly dependent on the system size, rather than on the number of gates to be processed.
We can notice negative speedups in the left half of the heatmap, as the \textit{qsim-cuda} backend outperforms the two other simulators. On the second half of the heatmap, we can appreciate noticeable speedups on both backends. In the QAOA, Random, QPE and QFT circuits the speedups range to up to $60\times$ for the \textit{cusv} backend and up to $53\times$ for the \textit{cutn} backend. We can notice larger speedups on the Hidden Shift, Hamiltonian Simulation, VQE and Bernstein-Vazirani circuits, which exceed the $5000\times$ speedup value for Hamiltonian Simulation and the VQE. These last four circuits are the ones that scale more slowly in terms of the number of quantum gates, as previously seen in Figure \ref{scaling_number_gates}. The main outlier is the Random circuit, which has poor performance on the \textit{cutn} backend, with a negative speedup at high qubit sizes.

\subsection{Distributed sliced tensor contraction performance}
In order to understand how the performance of tensor network contraction scales as we increase the number of GPUs, we designed a strong scaling experiment.
We implemented a distributed version of the tensor network contraction algorithm, by leveraging the cuTensorNet library, MPI and NCCL, and ran scaling experiments for all of the circuits in the benchmark at size 32 qubits, apart from the Random circuit which was limited at size 28 qubits,
by using an increasing number of GPUs and compute nodes on the Leonardo Supercomputer, with the objective to test the efficacy of tensor network slicing in improving contraction efficacy.
The algorithm starts by spawning one MPI process for each available GPU, and first performs a distributed pathfinding on the whole network. The best path, selected according to the lowest FLOPs count, is broadcast to all other MPI processes thorugh the MPI communicator.
The tensor network is then sliced in a number of sub-networks equal to the number of MPI processes, in order to provide to each MPI process, and thus each GPU, a comparable amount of FLOPS to be performed. Each GPU contracts its own sub-network, and all the partial results are reduced with a sum operation through the NCCL communicator, that yields the final amplitude result.

\begin{figure}[!ht]
    \centering
    \includegraphics[width=\linewidth]{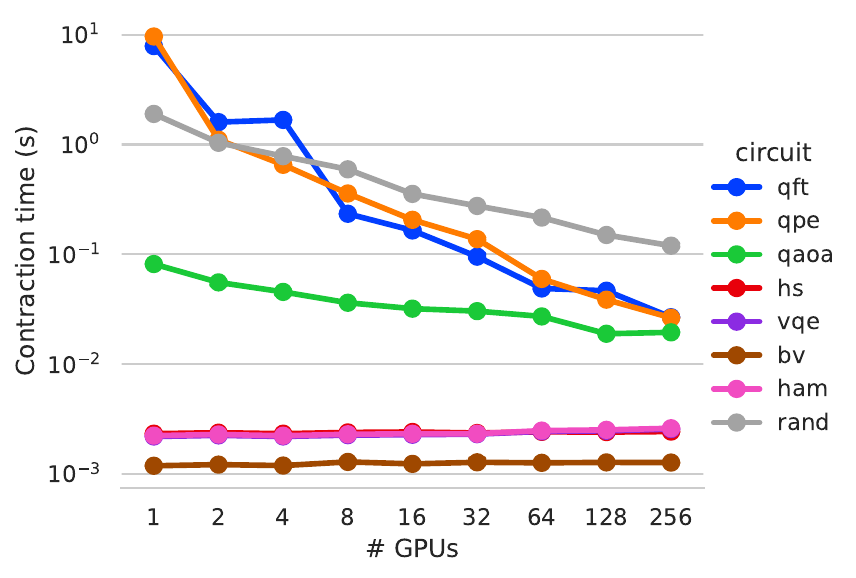}
    \caption{Strong scaling of distributed sliced tensor network contraction performance on circuits of size 32 qubits, with the exclusion of the Random circuit at size 28 qubits. The number of GPUs (\textit{\# GPUs}) also corresponds the number of MPI processes. Each point represents the mean of 30 datapoints.}
    \label{sliced_strong_scaling}
\end{figure}

\begin{figure*}[!ht]
    \centering
    \includegraphics[width=1\linewidth]{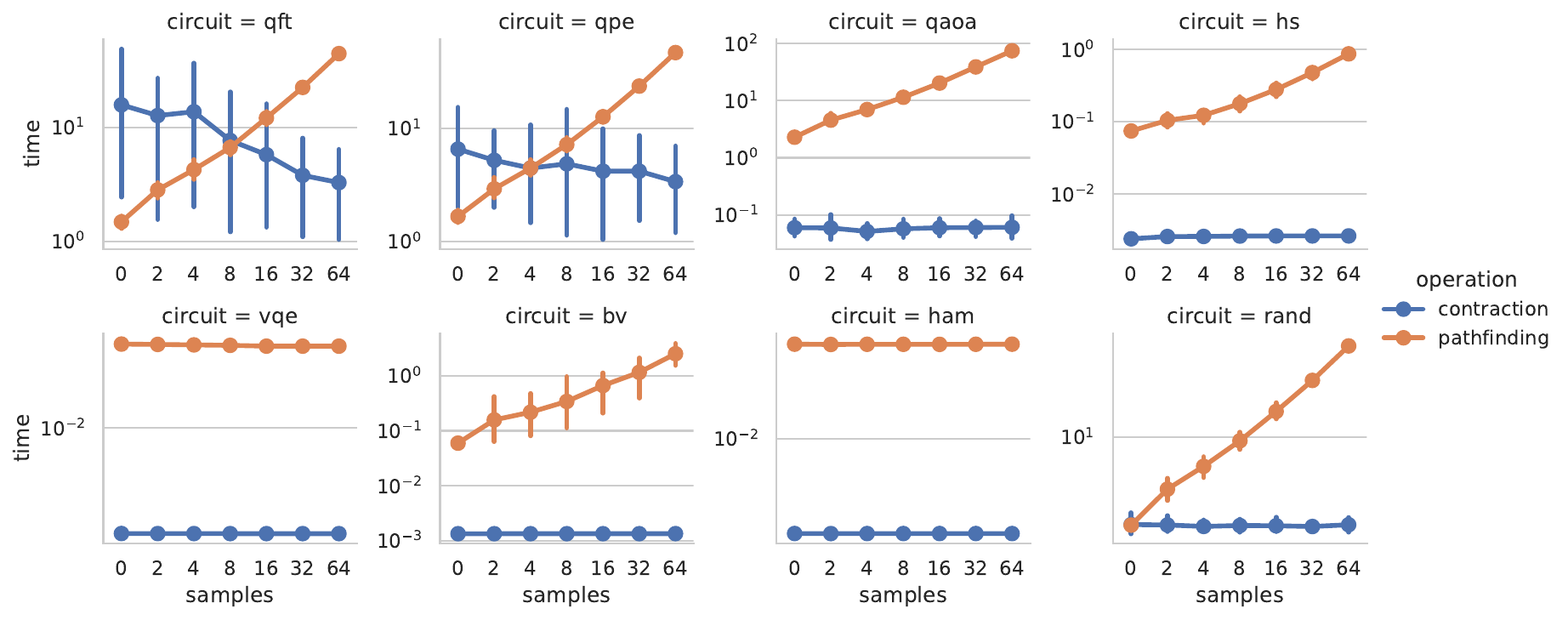}
    \caption{Pathfinding time (orange) and contract time (blue) for tensor networks derived from circuits of size 32 qubits. The number of samples performed by the pathfinder are executed sequentially in order to extrapolate the total time required for this computation, whilst contractions are performed using a single GPU. Notice different y-axis limits for each circuit in the benchmark. Vertical bars represent the 90th percentile interval.}
    \label{pathfinding_vs_contract}
\end{figure*}

In Figure \ref{sliced_strong_scaling}, we present the strong scaling results of this experiment, from using 1 GPU going up to 256 GPUs. The node configuration on Leonardo includes 4 GPUs per node, and each node is interconnected with an Nvidia Mellanox network, reaching up to 200 Gbit/s node to node transfer rates. Given the size of the partial results being reduced, network bandwidth does not act as a computational bottelenck.
We can see from the strong scaling results that not all the quantum circuits selected in the benchmark exhibit large performance gains, particularly the Hidden Shift, VQE, Bernstein-Vazirani and Hamiltonian simulation. These circuits lack complexity in the structure of their circuit-derived tensor networks, as reported by their asymptotically decreasing program communication metric in relation to the size of the quantum circuit.
The QAOA sees a noticeable improvement in terms of contraction performance, steadily lowering the contraction time from 81 ms on a single GPU to as low as 19 ms on 16 GPUs, a speedup of more than $4.2\times$.
Likewise, the Random circuit's contraction time goes from a mean of $1.89$ s when using a single GPU to about 120 ms in the case with 256 GPUs, a speedup of about $15\times$. 
The largest improvements in the contraction times can be obtained on the QFT and QPE quantum circuits, which drop more than one order of magnitude in contraction time when going from running on one GPU to 16 GPUs: we respectively measure a speedup of more than $294\times$ on the QFT circuit and a speedup of more than $364\times$ on the QPE circuit, in lieu of just increasing by $256\times$ the computational resources available.

\obs{
Quantum circuits with large program communication scores benefit the most from sliced distributed contraction, reaching superlinear speedups with respect to a linear increase in computational power.
}

Albeit some of the benchmark circuits considered have been found to be trivially solvable even by a single GPU, we demonstrated how specific descriptive metrics of a quantum circuit, namely the program communication metric, can let us foresee the presence or absence of a computational gain with a distributed sliced tensor network contraction.

On the topic of evaluating distributed sliced performance, it is still unclear how to measure weak scaling performance of different quantum circuits, that is measuring the performance of a problem when scaling equally the problem's complexity and the available computing resources. The problems considered in this benchmark are parameterised by the number of qubits, which does not provide direct control over the problem's contraction complexity, which mainly depends on the treewidth of the tensor network \cite{Gray2021}.
It could be possible to develop a synthetic parameterisable quantum circuit that grows in terms of the circuit derived tensor network's treewidth.
This would most probably end up being a variation of a Random circuit, which however holds no meaning in terms of problem solution.

\obs{
The complexity of contracting quantum circuit derived tensor networks does not scale with the problem size, i.e. the number of qubits. As such, specialised synthetic benchmarks are needed to measure the weak scaling of tensor network contraction. 
}

For the sake of this article, we can safely predict a correlation with the strong scaling capacity of the cuTensorNet library in relation to the program communication metric of a quantum circuit. For quantum circuits with program communication scores of one can efficiently leverage large multi-GPU acceleration. On the opposite case, when the program communication approaches zero, one GPU can suffice, and no advantage is gained from increasing computational resources.

\subsection{Pathfinding impact on tensor contraction performance}
\label{pathfinding_contraction_performance}
There is a need to classify quantum circuit derived tensor networks according to their pathfinding complexity. Specifically, we are interested in predicting which circuits can be contracted with higher efficiency in correlation to an increase in the resources available to the pathfinding algorithm.
We investigate how a variation in the resources available to the cuTensorNet pathfinding algorithm, namely the number of samples performed, impacts on the contraction time of the quantum circuit derived tensor networks in the benchmark.
We measure the total time required by the pathfinding algorithm to sequentially compute all of the samples by enforcing a single thread. Although this search may be easily distributed, the total computation time required by the pathfinder is still the same.
The size of each quantum circuit is kept fixed at 32 qubits, apart form the random circuit of size 28 qubits, due to representability reasons.

In Figure \ref{pathfinding_vs_contract}, we plot the variation in the pathfinding and contraction time with respect to an increase in the number of samples, and find three categorisations for the quantum circuits used in this benchmark: pathfinding-bound problems, contraction-bound problems and unbounded problems.
Pathfinding-bound problems include the VQE and Hamiltonian simulation circuits, since the complexity landscape of their possible contraction paths is mostly flat. This means that such circuits provide no advantage in terms of contraction time when assigning additional resources to the pathfinding algorithm.
Contraction-bound problems include the QAOA, Hidden Shift, Bernstein-Vazirani and Random circuits, and they are characterised by having a non-trivial pathfinding complexity landscape, but trivial contraction complexity. This means that providing additional resources to the pathfinder indeed gives rise to better solutions in terms of total FLOPS in contraction, but the overall difference in terms of contraction efficiency is unnoticeable.
Unbounded problems include The QFT and QPE problems, since they show an direct correlation between the amount of resources provided to the pathfinding algorithm and a reduction in the contraction time.
In fact, as the number of samples increases, in both cases we can measure a speedup in terms of contraction time of $1.9\times$ on the QPE and of $4.79\times$ on the QFT by increasing the pathfinding time by about $29\times$.
From these observations it follows that quantum circuits characterised by an \textit{entanglement ratio} metric that converges to \textit{one} map to more complex pathfinding problems that can efficiently leverage additional pathfinding resources, whilst if the same metric saturates to values lower than $0.4$, the problem is bounded, either in pathfinding or contraction. 

\obs{
Quantum circuit derived tensor networks can be classified in pathfinding-bound problems, contract-bound problems, and unbounded problems. Only the latter show an anticorrelation between pathfinding time and contraction time.
}

Although this result might point towards diminishing returns, it must be noted that the pathfinding time was purposefully measured in a sequential manner on a single thread, but each sample is indeed independent from the other, as such using one independent thread and core per sample is going to keep the real-time pathfinding time constant. Moreover, in terms of real-world tensor contraction, contractions are performed with a batched approach, where the contraction path is computed only once, but the actual contraction may be performed hundreds or thousands of times, thus outweighing the steep inital pathfinding cost.
Given the intense computational cost of pathfinding and its reliance on performing numerous samples, an open question remains on whether this algorithm can be further accelerated using GPUs in order to increase the number of concurrent samples while still fitting a low time bound.

\subsection{Lessons learnt}
% Flow chart for picking a simulator
% topological limits of the pathfinder
% An increase in the TN size significantly hinders the pathfinding performance

The simulation performance is ruled by a plethora of factors.
State vector simulators are inherently limited by the problem size, given the exponential memory requirement, thus trying to to simulate systems larger than $\sim50$ qubits with state vectors suffers from diminishing returns\cite{simulations_on_summit}. Moreover, the distribution of the subvector-matrix multiplications scales exponentially over the problem size, hindering the time performance. The main advantage lies in being able to access the complete set of information encoded in the wave function. Modern state vector simulators can take advantage of GPU caching in order to limit the computation time in small scale problems, but still hit a hard  memory bandwidth limit as soon as the cache's capacity is saturated.
Tensor Network contraction has a lot of potential for problems which embody a structure with well balanced connectivity, as that of a structured mesh, as this generates small sized intermediate tensors during the contraction process. As such, current pathfinding algorithms are optimised for finding such structures, and struggle whenever the topology becomes more irregular, such as when most of the entanglement operations in a circuit are concentrated on a few qubits. This is because the size of the intermediate tensors immediately spikes up, slowing the dot product with the remainder of tensors. The main advantage of this approach is the fact that it scales linearly in memory, thus allowing the simulation of larger systems, so long as the circuit's structure is favourable for contraction. In fact, by scaling the number of qubits in order to saturate the memory capacity of an 80GB NVIDIA A100 GPU, assuming that connectivity information is stored in the shared memory of the machine, one would need more than $7.5k$ logical qubits for a QAOA circuit with $P=1$, or more than $40k$ qubits for the Random circuit, the two circuits in the benchmark with the highest gate scaling per qubit. Despite the fact that there is no guarantee of convergence towards an optimal path, tensor network contraction is the only approach to exact simulation of quantum circuits that can scale favourably to circuits with dimensions larger than $50$ qubits. Previous works show that parallelising the contraction process is trivial, thus the main bottleneck remains the pathfinding algorithm, a well known NP-hard problem \cite{pednault2020paretoefficient}.

As we have observed, circuits which are characterised by an \textit{entanglement variance} score greater than $0.2$ have a highly unbalanced structure, reducing the efficacy of the pathfinding algorithm for tensor network contraction.
Likewise, if more than half of the circuit's gates are double qubit gates, which amounts to a \textit{entanglement ratio} score greater than $0.5$, the tensor network approach starts to scale poorly, as stated in Observation 4. This is due to the presence of large tensors early on during the contraction, slowing down the overall process.
In both of the previous cases is thus suggested to use a state vector solver.
However, if these first two conditions are not met, one must check for the \textit{program communication} and the \textit{critical depth} scores. If they are opposite to one another, with one being smaller than $0.2$ and the other being larger than $0.9$, then the best simulation method is the tensor network contraction. This is due to the fact that, if the previous conditions about the \textit{entanglement ratio} and the \textit{entanglement variance} scores are met, a \textit{program communication} score greater than $0.9$ and a \textit{critical depth} score lower than $0.2$ indicate a circuit structure in which most qubits interact with each other, but there are little to no repeated interactions. On the other hand, a \textit{program communication} score smaller than $0.15$ together with a \textit{critical depth} score larger than $0.9$ indicate a circuit structure composed of many chained two-qubit interactions among the same pairs of qubits. In both cases the tensor network becomes a pseudo-regular grid, which can be efficiently contracted whilst keeping intermediate tensor sizes at bay, leading to time performance gains of up to one order of magnitude.

\textit{Program communication} is especially important in determining whether a quantum circuit contraction can be efficiently contracted in a distributed sliced setting. Those circuits display superlinear speedups with respect to the available compute resources.
Moreover, we showed how is is only justifiable to provide additional pathfinding resources to unbounded quantum circuit tensor contractions, as they can provide further speedups, whilst all other circuits can save on using additional resources on pathfinding.

\section{Conclusions}
\label{conslusions_sec}
% What did the paper contribute? Sum up the results
% Is there room for further improvement or further study?
The paper characterised the performance of a selected suite of relevant quantum circuits when simulated on state of the art simulator backends and high performance hardware. At first, the circuits have been characterised according to objective metrics that could describe their topological structure, to later correlate them to the performance of the simulators. The results point towards the fact that statevector simulators become heavily communication bound as soon as the size of the statevector exceeds that of the GPU's cache. The tensor network contraction has proven to perform better in circuits that have well distributed entanglement among the qubits, with a low overall number of two qubit gates in relation to the total number of gates.

It is reasonable to assume that by improving the memory access pattern of a state vector backend, one may improve the time performance of any benchmark to be simulated.
The tensor network backend proved to be highly dependent on the efficacy of the pathfinding algorithm.
The overall performance of this second backend can already outpace the state vector simulator in some of the benchmark problems, whilst keeping a comparable, albeit slower performance in the remainder of the test runs.
It is reasonable to assume that the development of further pathfinding optimisations for these harder to tackle topologies may push the performance of the tensor contraction backend to become the fastest simulation approach for quantum circuits, for example by limiting the growth of intermediate tensors during contraction.
This would let us leverage the fact that the representation of tensor networks in memory enables the validation of larger quantum circuits and computers.
Moreover, the promising scaling of distributed pathfinding and sliced contraction approaches over large tensor networks have the chance to further close the gap on simulating real quantum computer. 
We highlight the absence of a class of real-world quantum algorithms with parameterisable contraction complexity. The development of such a class of algorithms would provide means for measuring weak scaling performance of distributed tensor network contraction libraries, easing performance comparisons between quantum and classical systems.
Future works could investigate the applicability of GPU accelerated pathfinding algorithms, so as to further improve the path quality and reduce the overall contraction time in unbounded problems.

%%%%%%% -- PAPER CONTENT ENDS -- %%%%%%%%

%%%%%%%%% -- BIB STYLE AND FILE -- %%%%%%%%
\bibliographystyle{IEEEtran}
\bibliography{refs}
%%%%%%%%%%%%%%%%%%%%%%%%%%%%%%%%%%%%

\end{document}